\documentclass[fleqn,usenatbib]{mnras}

\usepackage{mathptmx}
\usepackage[T1]{fontenc}

\usepackage{graphicx}	
\usepackage{amsmath}	
\usepackage{amssymb}	

\title[Quasar ID confusion]{Identification confusion and blending concealment in the SDSS-DR16 Quasar catalogues -- 40 new quasars and 82 false quasars identified.}

\author[E. Flesch]{\textbf{Eric Wim Flesch }$^{1}$\thanks{E-mail: eric@flesch.org}
\\
$^{1}$PO Box 15, Dannevirke 4942, New Zealand}

\date{Accepted XXX. Received YYY; in original form ZZZ}

\pubyear{2021}

\begin{document}
\label{firstpage}
\pagerange{\pageref{firstpage}--\pageref{lastpage}}
\maketitle

\begin{abstract}
The SDSS-DR16 Quasar Superset and pipeline catalogues are searched for undeclared quasars which were concealed by or confused with other objects, usually due to incomplete deblending.  Forty such quasars with redshifts are found and herewith presented.  Also, 82 entries in the SDSS-DR16Q main quasar catalogue are shown to be non-quasars, some also due to incomplete deblending, especially ``line poachers''.  Other non-quasars are shown to be star spikes or moving stars or other types, and 7 new quasars are identified from analysis of ``unplugged'' spectra.  These numbers are tiny compared with the millions of DR16 spectra, but highlight some heretofore overlooked systematics.  The deblending issue is discussed throughout.     
\end{abstract}



\section{Introduction}
The Sloan Digital Sky Survey (SDSS) Quasar Catalogue 16th Data Release \citep[DR16Q:][]{DR16Q} consists of two files, being the quasar-only main catalogue of 750\,414 entries which includes earlier SDSS-I/II/III visually confirmed quasars, and a 1\,440\,615 row ``superset'' of SDSS-IV/eBOSS quasar target classifications.  The SDSS-DR16 pipeline catalogue \citep[DR16:][]{DR16} gives pipeline-processed details of 5\,789\,200 spectra taken over the life of the SDSS project, including multiple spectra for some targets.  Data is taken from each of these for this work.

The DR16Q uses quite a different methodological approach to data processing \& presentation compared with its predecessor DR14Q \citep{DR14Q}; it emphasizes data uniformity and rule-based classifications whereas DR14Q (and its predecessors) had a more heuristic approach.  This rigour produces quantifiably reliable data for onwards research but can overlook data anomalies which evade the data-gathering rules.  I found such anomalies in the overall SDSS data which yielded 40 evident quasars which had not heretofore been identified as quasars.  Such identifications need redshifts to be useful, and those redshifts are available as well: the DR16Q catalogues present multiple redshifts per object as are available, including the neural automated QuasarNET \citep{QuasarNET} redshift for which is claimed $>$99\% efficiency and $>$99\% accuracy within 0.05z.  I report QuasarNET redshifts and DR16 pipeline redshifts in this work.  

The 40 new quasars come in groups depending on the method to collect them, plus there is some history of how one group led to another, so I present them in groups accordingly.  At the outset, I show some anomalous DR16Q data to introduce the topic of things not always being what they seem, and document the ``2 arcsecond rule'' which can confuse quasars with their neighbours on the sky.  Other quasars were concealed in the merged photometry of close doublets or by the glare of bright stars.  And a technique of swapping out ``unplugged'' objects is detailed, with 7 quasars added by that.  There are 22 figures in this paper, some quite reduced in size, but just zoom in to see them clearly.  

At the end, I present 82 DR16Q ``quasars'' which are clearly not quasars, such as, for example, if they should be star spikes.  In such a large 750K data pool there is no great significance in these 82 non-quasars except to reify SDSS-given caveats which were stated only statistically, and to introduce the topic of ``line poaching'' which can cause non-quasars to be visually classified as quasars.  The topic of spectral contamination by the emission of close neighbours on the sky is a repeated theme throughout this paper.

\section{Anomalous data and the 2-arcsecond rule}

The DR16Q paper abstract states an expectation of ``0.3\%-1.3\% contamination'' in the quasar-only catalogue.  Figure 1 shows 3 such contaminants, star spikes in this instance.  

\begin{figure} 
\includegraphics[scale=0.38, angle=0]{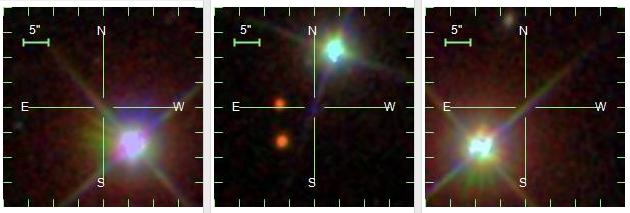} 
\caption{DR16Q: ``visually inspected'' quasar entries seen to be star spikes at J154253.66+351540.1, J155312.39+442233.3, and J172627.56+352447.9 respectively.} 
\end{figure}

What is instructive about those 3 examples is that they were visually inspected (so flagged) and assigned redshifts $>$2 with medium or high confidence.  This shows us that ``visual inspection'' consists of inspection of the spectrum but not of the image.  This lack of the image as a resource opens the door to artefacts of what can be called ``the 2-arcsecond rule'' which is given as\footnote{at http://www.sdss.org/dr16/spectro/catalogs.php} ``\textit{Any group of spectra which are within 2 arcsec of each other are considered to be of the same object, and only one of them is designated as `sciencePrimary'}".  This rule prevented spurious duplicates, but the co-mingling of spectra plus spectral deblending limitations and the absence of the image could cause close objects to become confused, as with the following example.  

\begin{figure} 
\includegraphics[scale=0.36, angle=0]{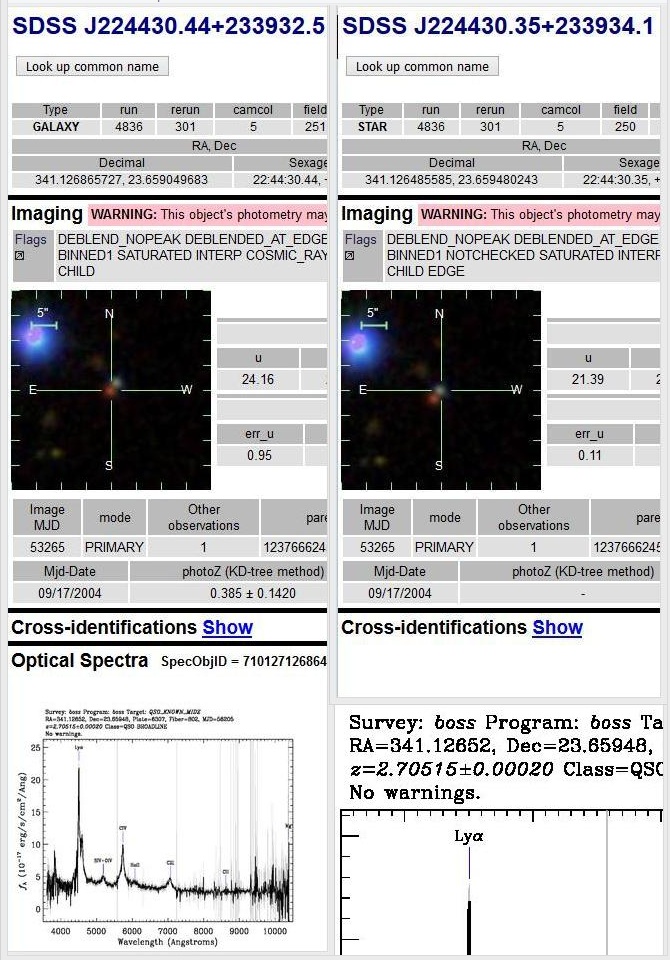} 
\caption{A red-blue doublet with the spectrum wrongly attributed to the red object (left side).  The spectrum close-up at bottom right shows that the spectrum RA-Dec location is that of the blue object (upper right), not the red.} 
\end{figure}

Figure 2 shows SDSS finding charts\footnote{http://skyserver.sdss.org/dr16/en/tools/chart/chartinfo.aspx} for both objects of a red-blue doublet at J224430.44+233932.5, separated by 1.993 arcseconds.  The left half of the figure shows that a quasar-like spectrum is assigned to the red object.  The upper right of the figure shows that the blue object has no spectrum assigned to it.  The bottom right of the figure shows a close-up of the spectrum to display its decimal RA/Dec co-ordinates.  If you compare that to the decimal location of each object (seen above the word ``Imaging''), you will see that the spectrum is actually that of the blue object, not the red.  This spectrum is from the SDSS-III BOSS survey, and there is one other, a previous spectrum from the legacy SDSS-II survey which similarly targeted the blue object only.  So in fact the red object never had a spectrum taken at all.  

So how did this mis-linkage happen?  The legacy SDSS-DR7 \citep{DR7} pipeline catalogue and its contemporary SDSS-DR7Q \citep{DR7Q} visual catalogue assigned the legacy spectrum to the blue object, as did all subsequent SDSS pipeline editions to the present day.  However, subsequent visual catalogues did not report it again until the SDSS-DR12Q \citep{DR12Q}, which assigned the BOSS spectrum to the red object in spite of its target co-ordinates.  It's clearly an error and the only viable explanation is that the 2-arcsecond rule was built into the processing code and so caused the slippage.  This erroneous identification was passed onwards to the DR14Q and DR16Q catalogues.  As DR16Q does include the correct legacy DR7Q identification as well, the erroneous entry manifests as a close duplicate or as an apparent quasar doublet.

Another such example is a red-blue doublet at J135307.59+450854.1 with doublet width of 1.843 arcsec.  The situation is exactly the same but in this case DR7/DR7Q had neither object so the outcome is that DR16Q has the wrong identification only (again inherited from DR12Q/DR14Q) while the DR16 pipeline has the correct identification, as did its predecessor DR12 \citep{DR12} and DR14 \citep{DR14} pipelines.  Also, at the end of this paper, Figure 22 shows another case where the blue-object spectrum was seconded to the red object, again evidently due to the 2-arcsecond rule.  This confusion could have been avoided by resort to the images whereon the blue object would have been spotted at once, but we can use the images to unscramble doublets throughout this paper.   

\begin{figure} 
\includegraphics[scale=0.25, angle=0]{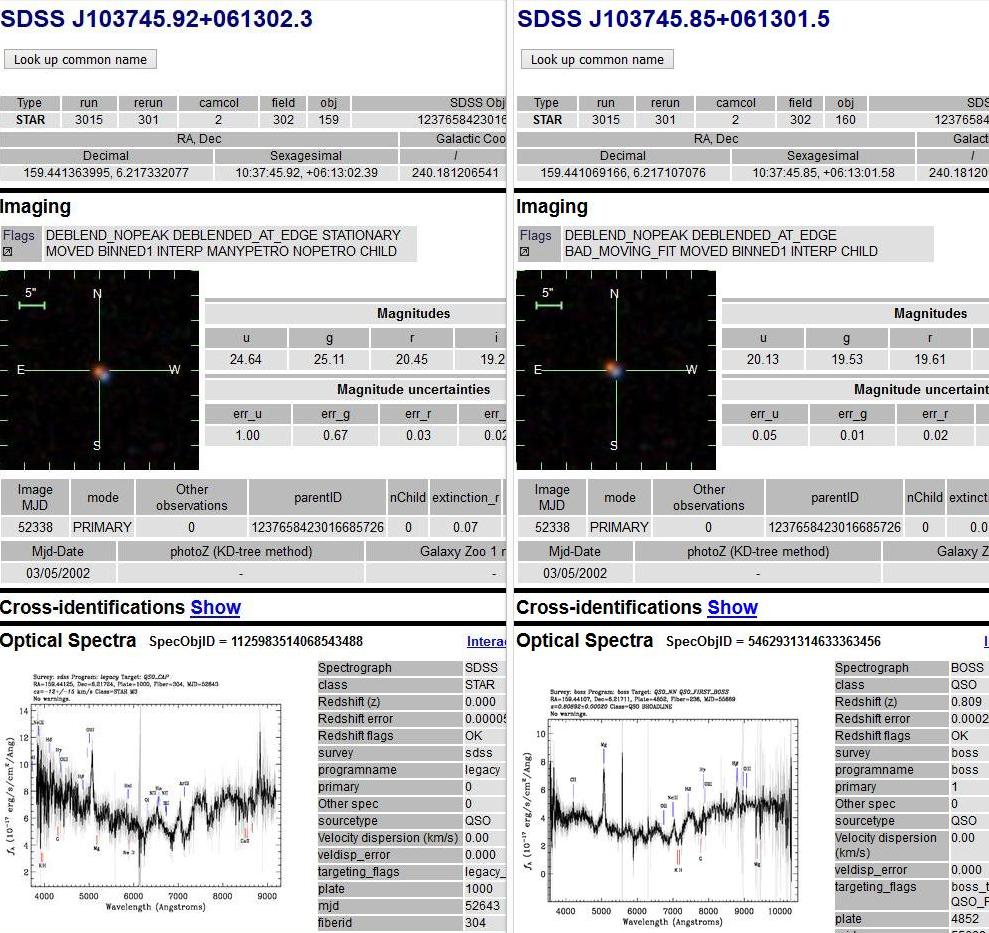} 
\caption{A red-blue doublet, both objects classified by DR16Q as quasars of z=0.809.  The blue object at right is pipeline-classified as a quasar, the red object at left is pipeline-classified as a star, but its spectrum was actually pointed at the joint object centroid, thus the two spectra look similar.  Separation between them is 1.270 arcsec.} 
\end{figure}

\begin{figure} 
\includegraphics[scale=0.35, angle=0]{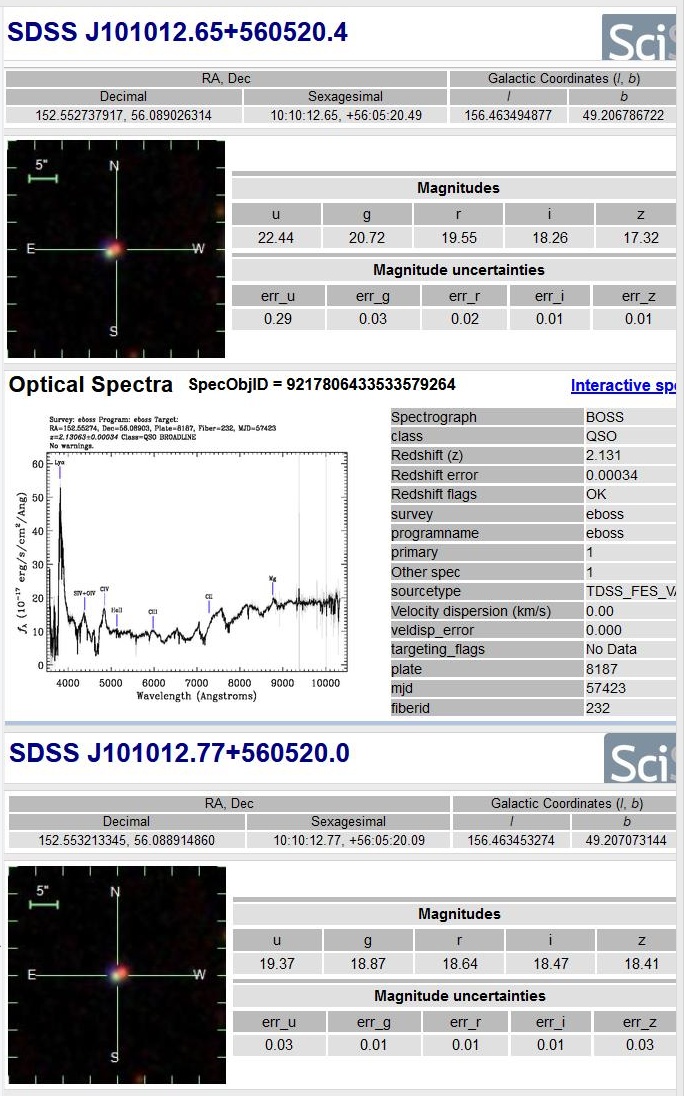} 
\caption{A red-blue doublet of separation 1.036 arcsec.  The spectrum is assigned to the red object (top), but the blue object (bottom) is the obvious source of the strong 3810\AA\ Ly$\alpha$ line because the red object is too faint in \textit{u} to encompass that line.} 
\end{figure}

\begin{figure} 
\includegraphics[scale=0.35, angle=0]{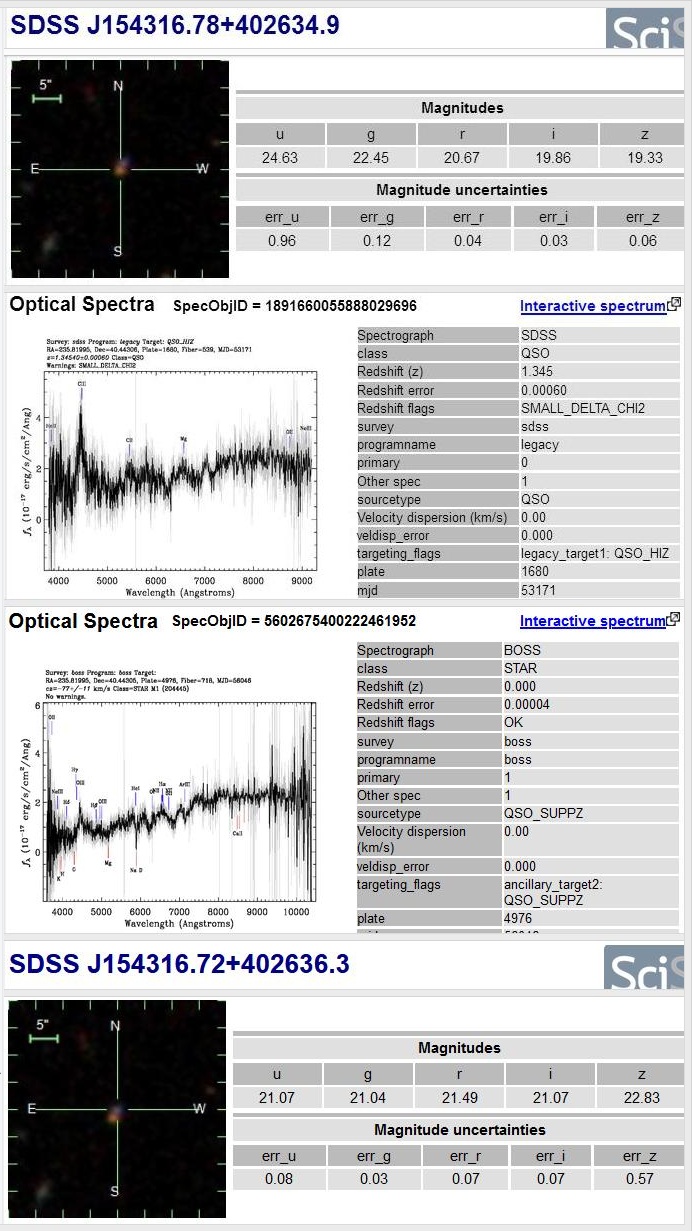} 
\caption{A red-blue doublet of separation 1.527 arcsec.  Both spectra show quasar lines, especially the upper one, and are linked to the red object (top) which however in the end was classified as an M1 star.  The blue object is thus the source of the quasar lines.} 
\end{figure}

\section{40 New Quasars identified within the SDSS-DR16 Quasar Superset and pipeline data}

DR16Q imported all quasar identifications from the legacy DR7Q catalogue, for completeness.  But a few of those had since been found to be not quasars and dropped, so DR16Q inadvertently restored those false objects.  As an example of this, Figure 3 shows SDSS-DR16 finding charts for both objects of a red-blue doublet separated by 1.27 arcseconds, thus within the ``2 arcsecond rule''.  The blue object on the right (J103745.85+061301.5) is a QSO of z=0.809.  The red object on the left (J103745.92+061302.3) has always been classified as a star by the pipeline, dating back to the legacy DR7.   However, DR7Q visually classified it as a QSO due to the emission lines contributed by the blue object -- the legacy spectrum\footnote{The 2-arcsecond rule has deprived the red object of a ``primary'' spectrum, note the information to the right of its spectrum which states ``primary''=0, i.e., this is not the primary spectrum, and ``Other spec''=0, i.e., there are no other spectra -- so, no primary spectrum.  Indeed, if you look up this object on the SDSS-DR16 finding charts it will display the object without any spectrum at all.  To get the spectrum, the user needs to select ''All Spectra'' to the left of the image, then the spectrum is made available.} (on the left) was actually of the joint object, the pointing being at the joint centroid.  A subsequent BOSS spectrum (on the right) was taken of the blue object instead of the red, allowing secure classification of the blue object as the quasar whilst dropping the red and so resolving the issue.  DR16Q includes the blue QSO but it also added the red ``QSO'' back in, from the DR7Q uptake.  Thus the error manifests as a spurious quasar doublet in DR16Q.

\begin{figure} 
\includegraphics[scale=0.35, angle=0]{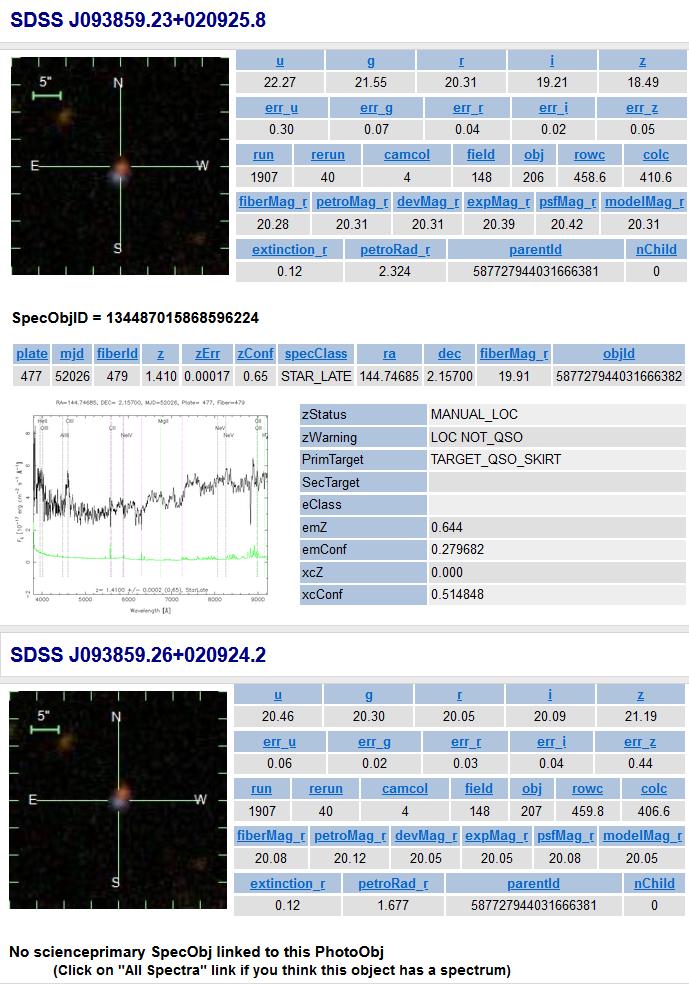} 
\caption{A red-blue doublet of separation 1.675 arcsec, from SDSS-DR7 finding charts.  The spectrum is of the joint objects but was assigned to the red object which was thus classified as a QSO in spite of the zWarning message (``LOC NOT\_QSO'') appended.} 
\end{figure}

\subsection{Three new quasars found in doublets with false legacy identifications}

With this as a guide, we can move on to 3 other red-blue doublets where the red object was wrongly decided by DR7Q/DR12Q to be a quasar, and so taken up by DR16Q.  The chief cause is imperfect deblending of the two objects which are so close that their spectra inevitably include emission from both objects.  This theme of spectral contamination from close neighbours recurs throughout this paper as it can cause stars to be identified as quasars, or conversely, quasars to be identified as stars.

Figure 4 shows a close red-blue doublet of separation 1.036 arcsec.  Both legacy spectrum and eBOSS spectrum (pictured) were onto the red object (J101012.65+560520.5) which has \textit{ugriz} colours typical of a red star.  The spectrum shows a red star continuum in the red wavelengths but BAL quasar lines in the blue wavelengths, including a bold 3810\AA\ Ly$\alpha$ line.  On the basis of the legacy spectrum, the legacy DR7 pipeline and DR7Q visual catalogues reported the red object as a quasar of z=2.13.  The subsequent DR10 \citep{DR10} \& DR12 pipeline catalogues switched the quasar identification to the blue object (J101012.77+560520.0) which has colours befitting a quasar, as an apparent manual intervention to select the better object.  However, the subsequent eBOSS spectrum was again onto the red object and again showed the quasar lines, so the DR14 \& DR16 pipeline catalogues reverted to identifying the red object as the quasar.  But no visual catalogue reported either object throughout this time.  DR16Q includes the red object as the quasar because it was taken up from DR7Q.  As the Figure 4 caption describes, only the blue object has a \textit{u} magnitude compatible with the strong Ly$\alpha$ line.  The blue object is the quasar, here presented as the first new QSO of this paper.  A complete listing of the 40 newly discovered quasars is given in Table 1.

\begin{figure}
\includegraphics[scale=0.35, angle=0]{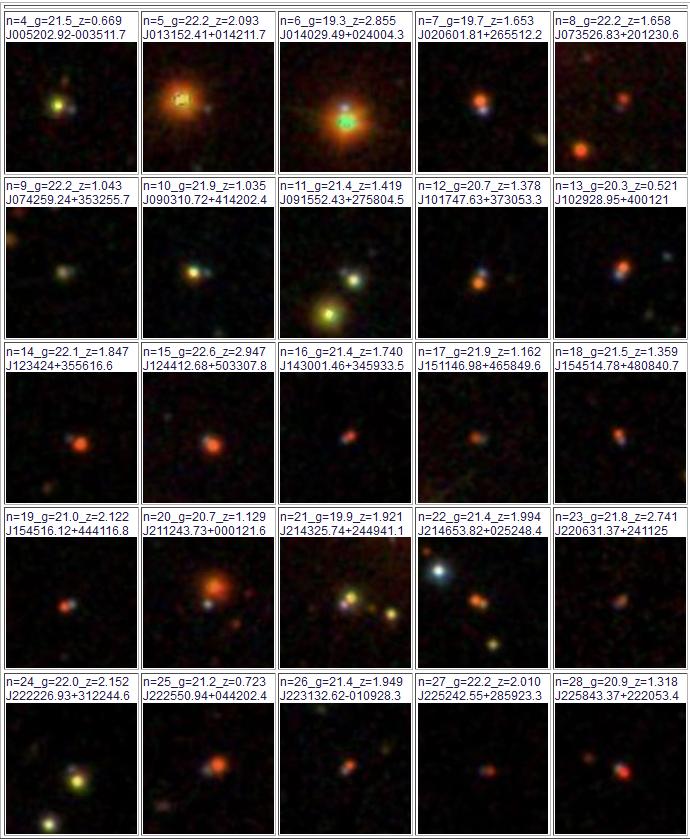} 
\caption{25 new QSOs previously concealed by their neighbours on the sky.  The quasars are at the exact centres of the images, which have 20 arcsec edges.  Image number (caption left) is the row number in Table 1, the caption also shows the SDSS-g magnitude and the QuasarNET redshift given by the DR16Q Superset which classifies all these as stars.  Figure 8 shows the spectra of these.} 
\end{figure}

Figure 5 is of a red-blue doublet of separation 1.527 arcsec, and features two spectra both of which are onto the red object (J154316.78+402634.9) which has \textit{ugriz} colours typical of a red star; the blue object (J154316.72+402636.3) has no spectrum and has flat photometry typical of a QSO of z$<$2.  The upper spectrum is legacy SDSS-II and shows a QSO of z=1.345, and the red object was thus so classified by the DR7 pipeline and DR7Q visual catalogues.  However, the subsequent lower spectrum (from BOSS) classified the red object as an M1 star; the difference is that the BOSS spectrum is better deblended from the blue object, but the broad 4500\AA\ emission line is still seen although less prominent.  Because of the BOSS spectrum, subsequent pipeline catalogues and the DR12 SuperSet catalogue all report the red object as an M1 star.  However, DR16Q includes the red object as a quasar because it was taken up from DR7Q.  The blue object, heretofore ignored, is obviously the source of the broad quasar lines and is the quasar, newly reported here.

\begin{figure*}
\includegraphics[scale=0.37, angle=0]{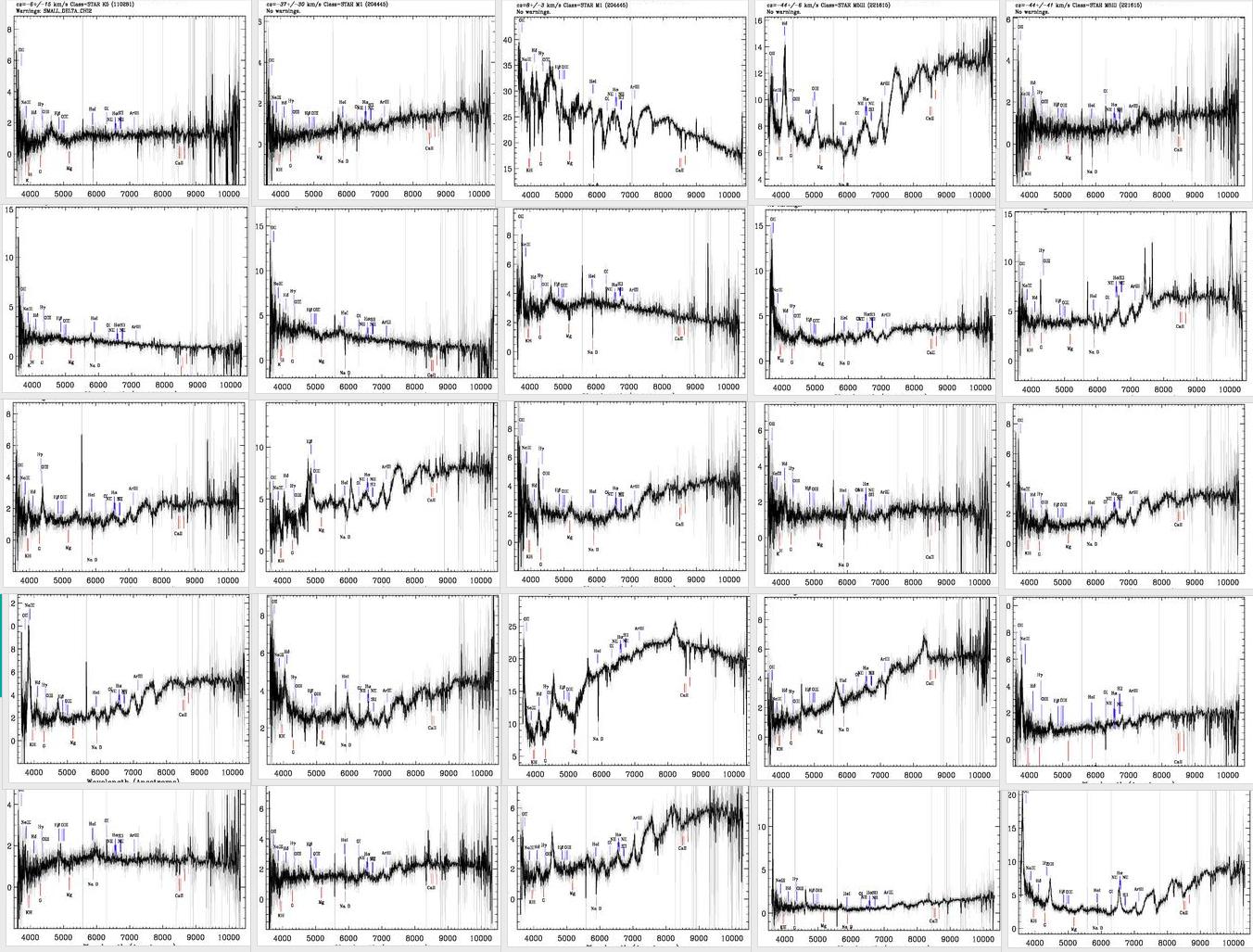} 
\caption{Spectra of the quasars of Figure 7, respectively.  All spectra are classified as stars by both the DR16 Pipeline and the DR16Q Superset due to stellar continua contributed by neighbours on the sky, but all show broad quasar emission lines.} 
\end{figure*}

Figure 6 is of a red-blue doublet of separation 1.675 arcsec; the SDSS-III photometric data reads it as a single object, so I use the legacy SDSS-II data to present the separate photometries.  The red object (J093859.23+020925.8) is assigned the spectrum shown and the \textit{ugriz} colours are typical of a red star; the blue object (J093859.26+020924.2) has no spectrum assigned and has flat photometry typical of a QSO of z$<$2.  The spectrum was pointed at the joint centroid so is of both objects but was assigned to the red object by DR7 because they were searching for high-redshift red quasars, and was classified as a QSO of z=1.410 by the DR7 pipeline and DR7Q visual catalogues.  However, subsequent SDSS-III BOSS processing of the same spectrum reclassified the red object as an M1 star, so later pipeline catalogues all report it as a star.  Nevertheless, DR16Q includes the red object as a quasar because it was taken up from DR7Q.  The blue object, heretofore ignored, is obviously the source of the broad quasar lines and is the quasar, newly reported here.

These total to 3 newly-discovered quasars previously concealed by being in close doublets with red stars; they are listed as the first 3 lines of Table 1.  This shows that a spectrum taken in the presence of a close neighbouring object can combine their emissions, even after efforts at deblending, such that the classifier can be confused into giving the classification of the neighbouring object instead of the targeted object.

\subsection{Twenty-Five Quasars concealed in doublets with stars} 
   
Here I present 25 quasars which were classified as stars because their spectra were contaminated by stray light from neighbouring stars on the sky.  Conversely, non-quasars can be classified as quasars when their spectra are contaminated by stray light from neighbouring true quasars on the sky; I will give 12 examples of that in Section 5. 

The DR16Q Superset presents all classifications of SDSS-IV/eBOSS quasar targets.  I searched its classified stars for star/QSO doublets which were photometrically dominated by the star, thus causing the pipeline to classify them as stars, and without visual inspection of the spectrum, and which showed quasar lines sufficient for the QuasarNET algorithm to classify them as quasars and give the redshift.  The rules used in the extraction were CLASS\_PERSON=0 (i.e., not visually inspected), AUTOCLASS\_PQN=``STAR'' (i.e., classified as a star), and IS\_QSO\_QN=1 (i.e., the QuasarNET algorithm classifies it as a quasar).  This yielded 1286 objects of which about half match to known quasars or pipeline quasars which are not new objects, with the remaining 634 being apparent stars.  I inspected those images and many spectra for visually convincing quasar lines, and found 25 to be QSOs concealed within star/QSO doublets, and 5 others described in the next section.  These are new QSOs, herewith presented. 
       
Figure 7 shows the 25 new quasars and Figure 8 shows their respective spectra.  All spectra were classified by DR16/DR16Q as stars but show quasar emission lines superposed onto stellar continua contributed by stray light from the neighbouring star.  QuasarNET redshifts for these are given by the DR16Q Superset and are shown on Figure 7 and listed in Table 1.

\begin{figure*} 
\includegraphics[scale=0.33333333, angle=0]{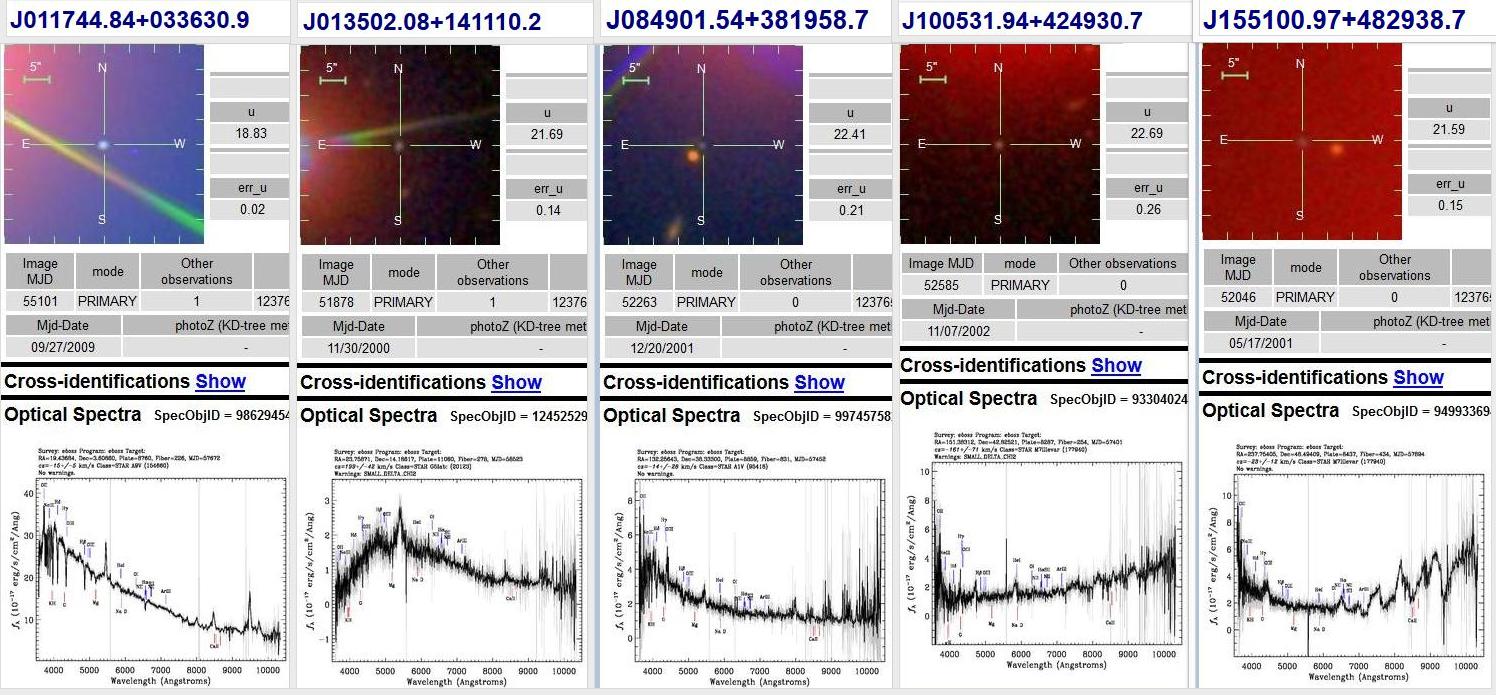} 
\caption{5 new quasars in the glare of bright stars.  Spectra show quasar lines superposed onto the stellar continua broadcast by the stars, causing these quasars to be classified as stars by DR16/DR16Q.  These quasars are listed respectively as rows 29-33 of Table 1.} 
\end{figure*}

\subsection{Five Quasars concealed in the glare of bright stars} 
	
The largest doublet separation of Figure 7 is 5.088 arcsec for the 2nd image, which suggests that bright stars can submerge quasar spectra at large offsets.  I have found 5 such new quasars, also from the DR16Q Superset extraction described in the previous section, see Figure 9 which shows images \& spectra.  The first one is a mag-18 quasar in the glare of 89 Pisces, a 5th magnitude star.  Table 1 gives their QuasarNET redshifts, plus more information about the bright stars.

\subsection{Seven Quasars salvaged from ``unplugged'' objects.}

The DR16 pipeline file gives a complete record of the spectral observations made on the Sloan 2.5m Telescope during the life of the SDSS project, which were 6826 plates taken with a total of 5\,789\,200 plugged CCD observations -- the first 2880 plates had 640 fibred plugs per plate, then 3946 later plates had 1000 plugs per plate.  Sometimes the fibres were not firmly plugged in, or even got their wires mixed up as to which CCD-plugholes they connected to, thus yielding a spectrum obviously incompatible with the photometry (such as a bright bluish spectrum for a faint reddish object).  SDSS documents such mix-ups with this caveat:\footnote{\scriptsize{at https://sdss.org/dr16/spectro/caveats/$\#$phot\%5Fspec\%5Fmatch}} ``\textit{With some frequency, the fiber mapping failed which identifies which fiber has been plugged into which hole. There are around 7200 such cases in DR10, which are marked as UNPLUGGED in the ZWARNING bitmask. The vast majority of these cases occur because the fiber was actually not plugged or was broken.  In around 200 cases, there is measurable signal down the fiber. In cases where there is more than one such fiber on plate, there is a possibility that the fiber location associated with the spectrum is incorrect (and thus that the photometric and spectroscopic information is mismatched)}''.  

\begin{figure}
\includegraphics[scale=0.2, angle=0]{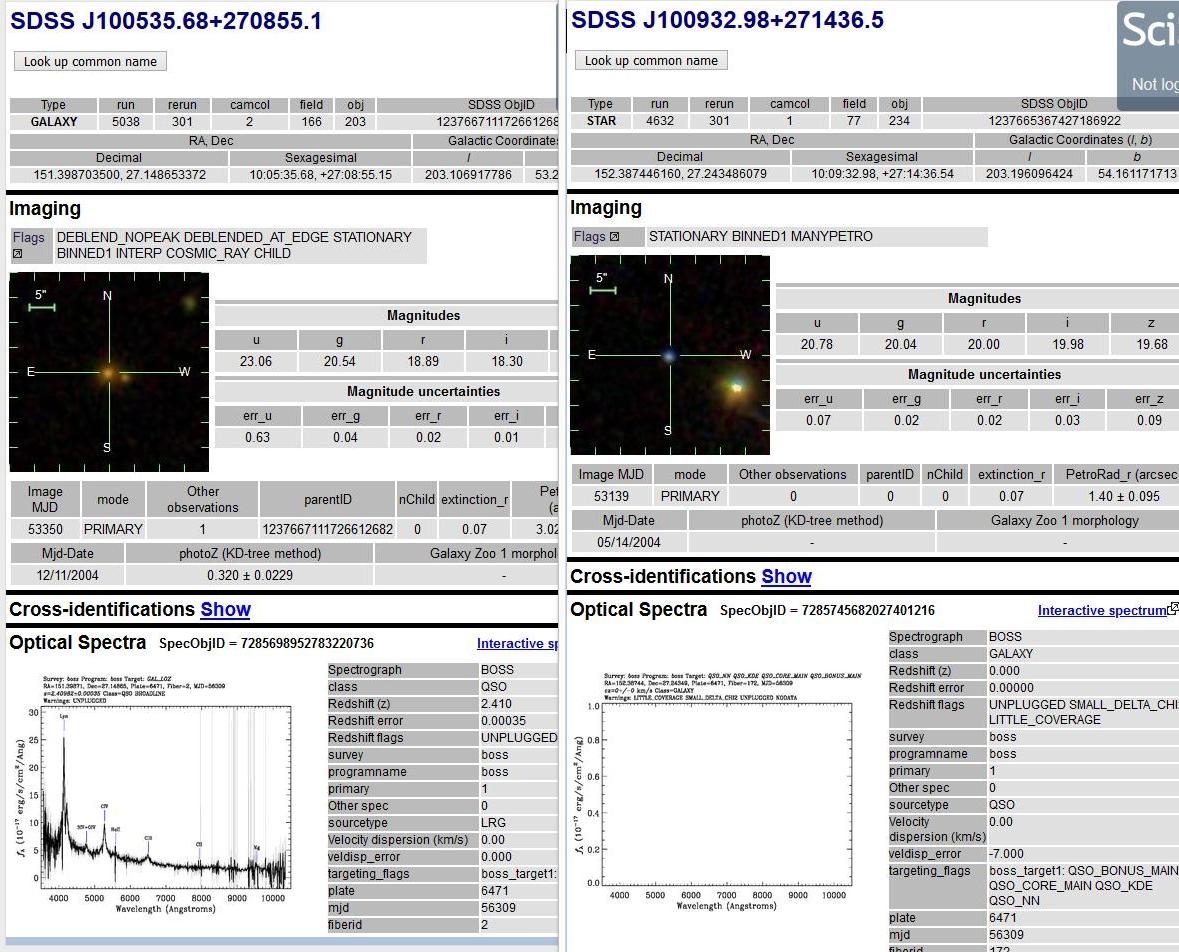} 
\caption{The only two UNPLUGGED fibres for plate 6471, MJD 56309.  Left side shows a quasar spectrum allocated to an unsuitable object (blueish spectrum, reddish object), right side shows an object with photometry well-matched to the quasar spectrum, and so, being the only candidate, adjudged to be the true source of that spectrum, and thus a quasar.} 
\end{figure}

\begin{figure*}
\includegraphics[scale=0.275, angle=0]{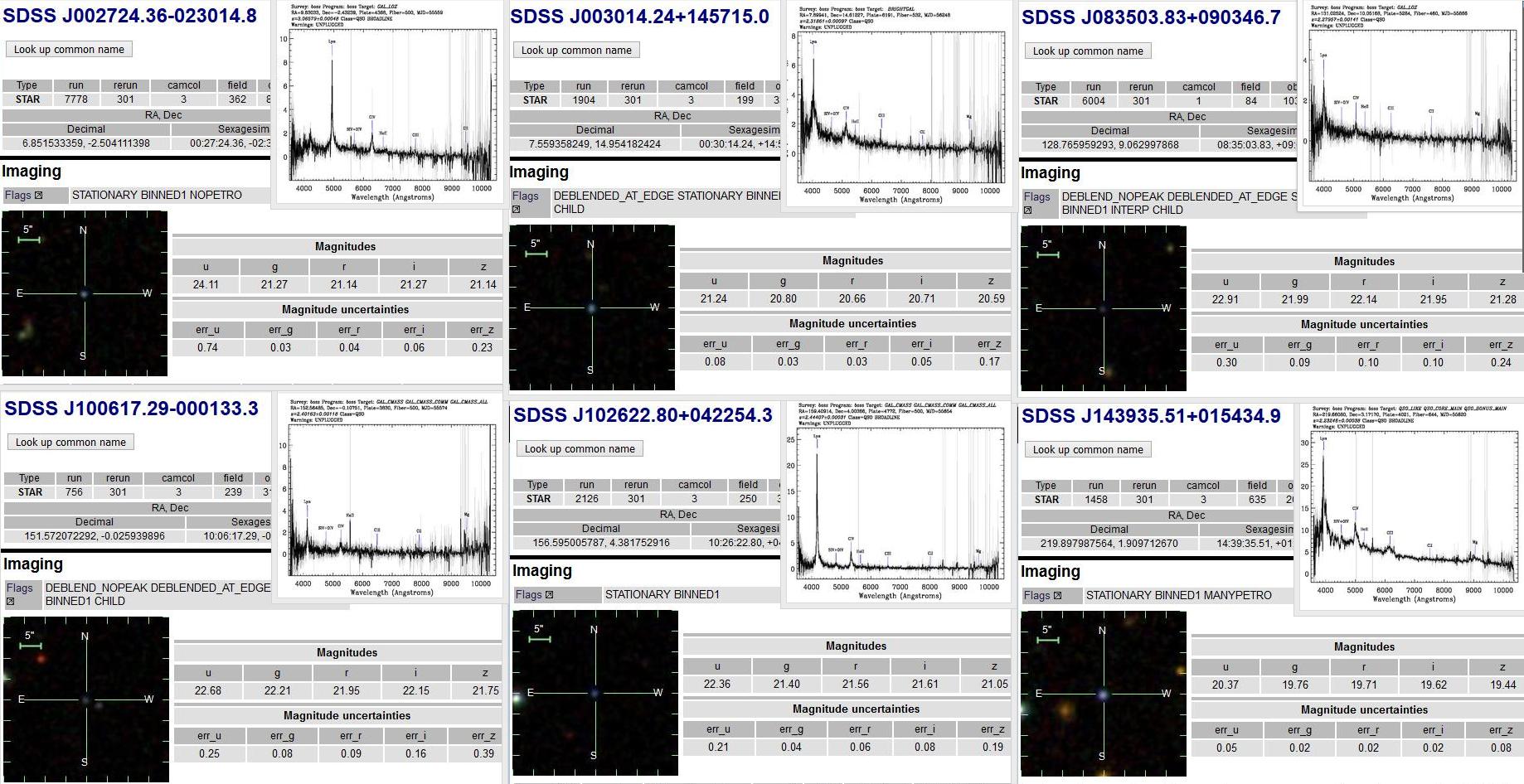} 
\caption{6 new quasars from UNPLUGGED swap-outs.  Each quasar has its spectrum on its upper-right, obtained from a fellow UNPLUGGED object (identified in Table 1) of the same plate-MJD combo.} 
\end{figure*}

\begin{figure}  
\includegraphics[scale=0.2, angle=0]{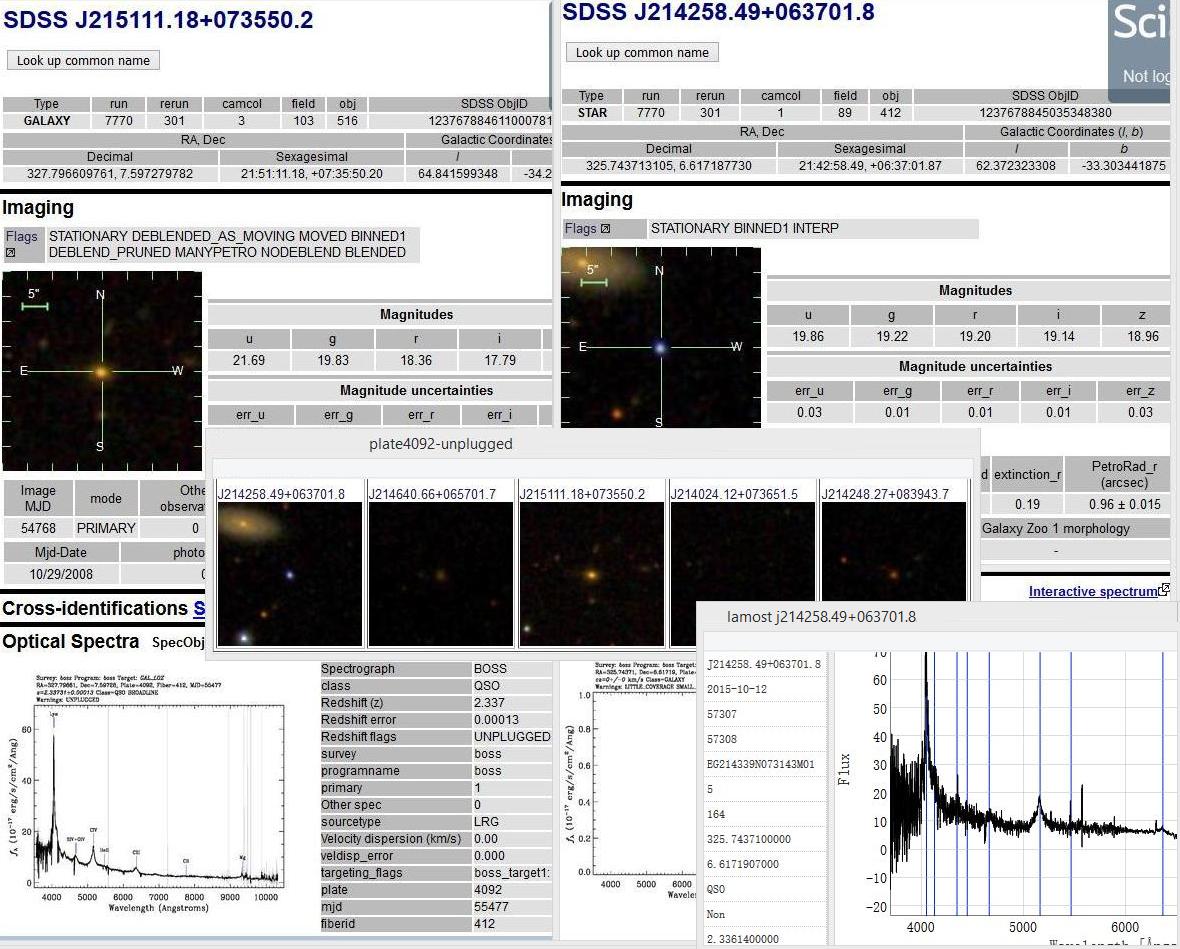} 
\caption{UNPLUGGED fibres for plate 4092, MJD 55477.  Left side shows the UNPLUGGED object wrongly allocated with the quasar spectrum (yellowish object, blueish spectrum), the band of 5 images superposed across the middle shows all 5 UNPLUGGED objects for this plate on which it is seen that only the left-most object has the right photometry (because it is blueish and of goodly flux) for the quasar spectrum, that object is identified at upper-right, and at bottom right is superposed a LAMOST-DR5 spectrum of that object which is seen to match the spectrum on the left, thus confirming the chosen object to be the quasar.} 
\end{figure}

\begin{table*} 
\scriptsize	 
\caption{The 40 new quasars}
\begin{tabular}{rllcccccccl} 
\hline 
\  & finding &                          &      &   z                &  \multicolumn{3}{c}{SDSS photometry} & doublet  & DR16 & \\
\# & method  & ID \& J2000              &   z  & source$^{\dagger}$ & \textit{u} & \textit{g} & \textit{r} & sep asec & zwarning$^{\star}$ & comment \\
\hline
1 & switch & SDSS J101012.77+560520.0 & 2.131 & pipe & 19.37 & 18.87 & 18.64 & 1.036 &   & replaces J101012.65+560520.4 as QSO, Sec 3.1  \\
2 & switch & SDSS J154316.72+402636.3 & 1.345 & pipe & 21.07 & 21.04 & 21.49 & 1.527 &   & replaces J154316.78+402634.9 as QSO, Sec 3.1  \\
3 & switch & SDSS J093859.26+020924.4 & 1.410 & DR16Q & 20.46 & 20.30 & 20.05 & 1.675 &   & replaces J093859.23+020925.8 as QSO, Sec 3.1  \\
4 & doublet & SDSS J005202.92-003511.7 & 0.669 & QNET & 21.60 & 21.53 & 21.73 & 2.547 & 4 & QSO classification was concealed by close star \\
5 & doublet & SDSS J013152.41+014211.7 & 2.093 & QNET & 22.37 & 22.18 & 21.97 & 5.088 & 0 & QSO classification was concealed by close star   \\
6 & doublet & SDSS J014029.49+024004.3 & 2.855 & QNET & 19.23 & 19.33 & 19.10 & 2.359 & 0 & QSO classification was concealed by close star   \\
7 & doublet & SDSS J020601.81+265512.2 & 1.653 & QNET & 19.83 & 19.72 & 19.58 & 1.576 & 0 & QSO classification was concealed by close star   \\
8 & doublet & SDSS J073526.83+201230.6 & 1.658 & QNET & 23.11 & 22.16 & 22.36 & 1.824 & 0 & QSO classification was concealed by close star   \\
9 & doublet & SDSS J074259.24+353255.7 & 1.043 & QNET & 22.23 & 22.24 & 22.08 & 1.421 & 0 & QSO classification was concealed by close star   \\
10 & doublet & SDSS J090310.72+414202.4 & 1.035 & QNET & 21.58 & 21.89 & 21.71 & 2.507 & 0 & QSO classification was concealed by close star   \\
11 & doublet & SDSS J091552.43+275804.5 & 1.419 & QNET & 21.44 & 21.40 & 21.23 & 2.108 & 0 & QSO classification was concealed by close star   \\
12 & doublet & SDSS J101747.63+373053.3 & 1.378 & QNET & 20.73 & 20.74 & 20.62 & 1.862 & 0 & QSO classification was concealed by close star  \\
13 & doublet & SDSS J102928.95+400121.0 & 0.521 & QNET & 20.62 & 20.28 & 20.28 & 1.311 & 0 & QSO classification was concealed by close star   \\
14 & doublet & SDSS J123424.00+355616.6 & 1.847 & QNET & 22.07 & 22.07 & 22.14 & 1.988 & 0 & QSO classification was concealed by close star   \\
15 & doublet & SDSS J124412.68+503307.8 & 2.947 & QNET & 22.24 & 22.60 & 24.41 & 1.443 & 0 & QSO classification was concealed by close star   \\
16 & doublet & SDSS J143001.46+345933.5 & 1.740 & QNET & 21.52 & 21.44 & 21.50 & 1.157 & 0 & QSO classification was concealed by close star   \\
17 & doublet & SDSS J151146.98+465849.6 & 1.162 & QNET & 21.48 & 21.94 & 21.82 & 1.264 & 0 & QSO classification was concealed by close star   \\
18 & doublet & SDSS J154514.78+480840.7 & 1.359 & QNET & 21.52 & 21.51 & 21.14 & 1.205 & 0 & QSO classification was concealed by close star   \\
19 & doublet & SDSS J154516.12+444116.8 & 2.122 & QNET & 21.09 & 21.01 & 20.88 & 1.272 & 0 & QSO classification was concealed by close star   \\
20 & doublet & SDSS J211243.73+000121.6 & 1.129 & QNET & 20.72 & 20.72 & 20.28 & 3.301 & 0 & QSO classification was concealed by close star   \\
21 & doublet & SDSS J214325.74+244941.1 & 1.921 & QNET & 19.88 & 19.88 & 19.58 & 1.571 & 0 & QSO classification was concealed by close star   \\
22 & doublet & SDSS J214653.82+025248.4 & 1.994 & QNET & 22.60 & 21.36 & 20.59 & 1.346 & 0 & QSO classification was concealed by close star   \\
23 & doublet & SDSS J220631.38+241125.0 & 2.741 & QNET & 21.89 & 21.82 & 21.71 & 1.084 & 4 & QSO classification was concealed by close star   \\
24 & doublet & SDSS J222226.93+312244.6 & 2.152 & QNET & 22.24 & 22.02 & 22.06 & 2.297 & 4 & QSO classification was concealed by close star   \\
25 & doublet & SDSS J222550.94+044202.4 & 0.723 & QNET & 21.66 & 21.17 & 21.03 & 2.017 & 0 & QSO classification was concealed by close star   \\
26 & doublet & SDSS J223132.62-010928.3 & 1.949 & QNET & 21.57 & 21.43 & 21.35 & 1.193 & 0 & QSO classification was concealed by close star   \\
27 & doublet & SDSS J225242.55+285923.3 & 2.010 & QNET & 22.02 & 22.19 & 22.44 & 1.143 & 0 & QSO classification was concealed by close star   \\
28 & doublet & SDSS J225843.37+222053.4 & 1.318 & QNET & 20.63 & 20.91 & 20.52 & 0.880 & 0 & QSO classification was concealed by close star   \\
29 & glare & SDSS J011744.84+033630.9 & 0.921 & QNET & 18.71 & 18.42 & 18.45 & 50 & 0 & in glare of 89 Pisces, a mag-5 blue star  \\
30 & glare & SDSS J013502.08+141110.2 & 0.934 & QNET & 21.74 & 21.18 & 21.33 & 25 & 4 & in glare of TYCHO 627-1089-1, a mag-10 star  \\
31 & glare & SDSS J084901.54+381958.7 & 1.843 & QNET & 22.16 & 22.21 & 21.75 & 45 & 0 & in glare of HD 75052, a mag-7 blue star  \\
32 & glare & SDSS J100531.94+424930.7 & 2.053 & QNET & 22.62 & 21.74 & 21.04 & 54 & 4 & in glare of NPM1+43.0405, a mag-10 red star  \\
33 & glare & SDSS J155100.97+482938.7 & 1.312 & QNET & 21.50 & 21.19 & 20.87 & 150 & 0 & in glare of HD 142143, a mag-7 red star  \\
34 & replug & SDSS J002724.36-023014.8 & 3.066 & pipe & 24.04 & 21.31 & 21.15 &   & 128 & spectrum swap from J003831.27-022556.6  \\
35 & replug & SDSS J003014.24+145715.0 & 2.319 & pipe & 21.16 & 20.85 & 20.67 &   & 128 & spectrum swap from J003135.84+143644.2  \\
36 & replug & SDSS J083503.83+090346.7 & 2.280 & pipe & 22.75 & 22.02 & 22.17 &   & 128 & spectrum swap from J084406.05+100306.0  \\
37 & replug & SDSS J100617.29-000133.3 & 2.402 & pipe & 22.58 & 22.24 & 21.97 &   & 128 & spectrum swap from J101015.56-000627.0  \\
38 & replug & SDSS J100932.98+271436.5 & 2.410 & pipe & 20.78 & 20.11 & 20.05 &   & 128 & spectrum swap from J100535.68+270855.1  \\
39 & replug & SDSS J102622.80+042254.3 & 2.444 & pipe & 22.29 & 21.45 & 21.56 &   & 128 & spectrum swap from J103738.19+040013.1  \\
40 & replug & SDSS J143935.51+015434.9 & 2.232 & pipe & 20.32 & 19.80 & 19.71 &   & 128 & spectrum swap from J143838.59+031018.1  \\ 
\hline
\multicolumn{10}{l}{$^{\dagger}$ redshift source: pipe=DR16 pipeline catalog, QNET=QuasarNET redshift provided by DR16Q Superset} \\
\multicolumn{10}{l}{$^{\star}$ ZWARNING: 0 = no warnings, 4 = pipeline classification not secure, 128 = unplugged / damaged fiber} \\
\end{tabular}
\end{table*}

That caveat isn't general enough, though.  Over 90\% of UNPLUGGED fibres have no signal at all, and certainly a mixed-up fibre-with-signal can match to a no-signal one, so it's sufficient for there to be just one UNPLUGGED fibre-with-signal for it to be matchable to any other UNPLUGGED fibre, whether signalled or not.  There are a total of 28\,248 UNPLUGGED fibres over all plates, thus an average of about 5 per plate for the 1000-plughole plates.  In the case of an UNPLUGGED but fully-signalled quasar spectrum, it would be desirable to determine which plughole sent that quasar signal.  However, SDSS production did not encompass such a deductive process -- the observation had to be done right for the data to be usable.  But given the data, sometimes it is clear which was the true source plughole.  Figure 10 shows a most obvious example, a plate which has only 2 fibres UNPLUGGED, with the left one being a reddish galaxy showing an incompatible blue quasar spectrum, and the other being a blue object with the right \textit{ugriz} profile for a z=2.4 quasar, as well as its magnitude being compatible with the spectral flux.  Therefore it is immediate that the blue object is the source of the quasar spectrum, and that quasar is one of the seven new quasars presented in this section.

Usually a plate has more UNPLUGGED objects than just 2, a typical example is plate 4092, MJD (modified Julian date) 55477, which has 5 UNPLUGGED objects one of which shows a quasar spectrum, see Figure 12 and its caption.  Only one of those objects has suitable photometry, so it is selected as being the true source of the quasar spectrum.  In this case we have a confirmation from LAMOST-DR5\footnote{http://dr5.lamost.org/v3} which targeted our selected object and presents a spectrum which matches our spectrum, see lower right corner of Figure 12.  This confirmation supports the validity of our technique, although this quasar is thus not newly-discovered.

While this may sound straightforward, in fact very few quasars can be identified in this way.  Only 562 UNPLUGGED objects have a quasar-classified spectrum and only about 100 of those are good-quality (2+ lines) spectra AND misplaced plugholes.  In many cases there are many candidate objects for the spectrum with no basis to choose one; sometimes one candidate looks much the best but that still is not good enough.  Sometimes a plate-MJD combo has multiple UNPLUGGED quasar spectra with which nothing can be done.  Only when there is just one good-quality UNPLUGGED quasar spectrum and just one viable candidate for it, can this method be used.  In the end I could make only 8 identifications, one of which was already known as shown in Figure 12.  Therefore 7 new quasars are included in Table 1 from this method, one of which was shown in Figure 10.  Figure 11 shows the remaining 6 new quasars, flagged as UNPLUGGED,  discovered from swapping out the spectrum with another UNPLUGGED object (identified in Table 1) from the same plate-MJD combo.  In every case this bluish object was the only eligible photometric match to a solo available quasar spectrum.  A notable point about these 7 swappers is that in 3 cases (1st, 4th, \& 5th spectra of Figure 11) the swapped fibre was $\#$500; perhaps it was the last one plugged with whatever remained.

Table 1 shows the full set of 40 new quasars presented in this paper.  The left column shows the method used to detect the quasar as described above: 3 switches within red/blue doublets where the blue QSO now replaces the red star as the identified QSO, 25 QSO-star doublets where the quasar spectrum was dominated by the stellar continuum contributed by the star, 5 QSOs previously washed out in the glare of bright nearby stars, and 7 UNPLUGGED quasars ``replugged'' by identifying the correct optical objects for those spectra.  The SDSS name (and thus J2000) is given, redshift and redshift provenance, some photometry and doublet separation where applicable, and a relevant comment.

\section{SDSS-DR16Q quasars which are not quasars}

It's not news that some SDSS-DR16Q objects classified as quasars are not in fact quasars, especially as the DR16Q paper abstract states an expectation of ``0.3\%-1.3\% contamination'' over its 750K quasars which thus indicates up to 10K non-quasars.  However, the bulk of those are expected to be within the low signal-to-noise faint objects where the spectra become more difficult to classify, that class of objects being retained for completeness.

What the DR16Q authors would not have intended to keep, though, would be obvious contaminants like star spikes and asteroids.  But some of those intrude nonetheless, and I have a few such to present here, maybe even the most of them, and the count is quite small compared with the very large 750K data pool, so for the record here are 3 asteroids mistaken for quasars, 38 star spikes mistaken as sources by the SDSS photometric reduction pipeline and subsequently accepted into DR16Q, 5 star glow reduction artefacts similarly accepted, 11 moving stars, 12 ``line poachers'' (as I call them) which show quasar emission lines only because of emission bleed-over from a neighbouring quasar on the sky, 6 supplanted objects from Sections 2 \& 3 above, and 5 processing duplicates.  Most of these are visually obvious so can be presented without elaboration, but I will explain the moving stars and processing duplicates, and the line poachers will require detailed justification in a separate section.  Table 2 lists all the non-quasars.

\subsection{Obvious non-quasars}

The next three figures show DR16Q-presented quasar entries which are seen to be obviously not quasars just by viewing their images.  Figure 13 shows 3 asteroids, Figure 14 shows 35 star spikes (additional to the three of Figure 1), and Figure 15 shows 5 star glow artefacts.  All have explanatory captions and show J2000 positions.  

\begin{figure} 
\includegraphics[scale=0.5, angle=0]{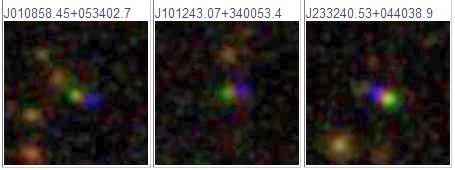} 
\caption{Three asteroids in DR16Q, the different colours alternating as the moving object crosses the field of view during the successive \textit{ugriz} exposures.  Images have 20 arcsec edges.} 
\end{figure}

\begin{figure} 
\includegraphics[scale=0.38, angle=0]{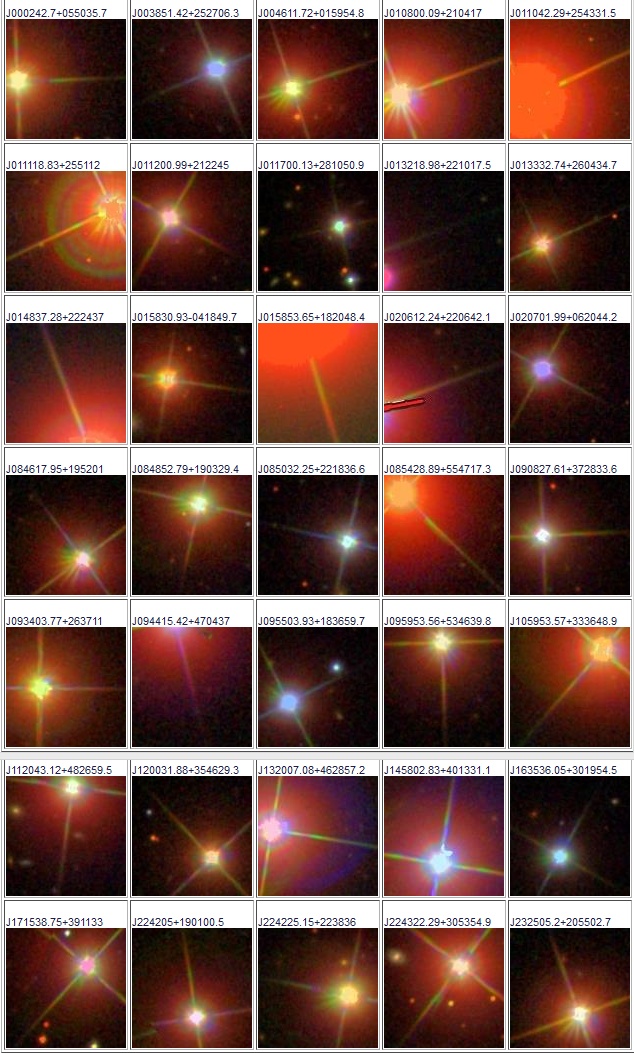} 
\caption{35 star spikes in the DR16Q main quasar catalogue, images have 1 arcmin edges.  Annotated DR16Q location is at exact centre of image, crossed by a star spike there with no background object present.} 
\end{figure}
     
\begin{figure} 
\includegraphics[scale=0.35, angle=0]{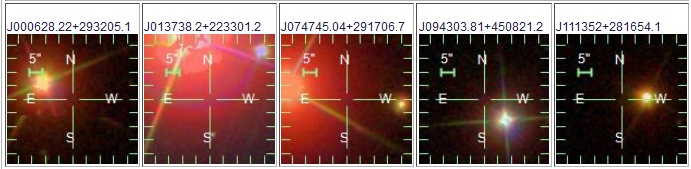} 
\caption{5 star glow artefacts in DR16Q, cross-hairs show there is no background object present at the annotated DR16Q location.} 
\end{figure}

\subsection{Moving Stars}

Eleven moving stars are listed in Table 2, rows 53 to 63; although classified as quasars in DR16Q, they have proper motion $>$25 mas/yr detailed in Gaia-DR2 \citep{Gaia} and confirmed\footnote{Confirmation of proper motion is needed because the moving star will necessarily become offset from the SDSS epoch to the Gaia epoch -- so near objects must be taken as candidates and then independently confirmed.} by inspection of epoch-1950s POSS-I images on DSS\footnote{DSS: the STScI Digitized Sky Survey at https://stdatu.stsci.edu/cgi-bin/dss\_form} which shows how far they have moved in the $\approx$50-year interval to the epoch-2000s SDSS images; the movements seen accord well to the Gaia-DR2 measurements as shown by the two examples on Figure 16.  All these 11 objects show spectra without clear quasar lines.  Two additional moving stars, J090952.54-014120.1 and J022723.99-010623.4, show strong quasar lines and are in fact co-positioned with background quasars in the SDSS epoch; those background quasars are faintly visible on the earlier-epoch DSS images at which time the stars were more offset.        

\begin{figure} 
\includegraphics[scale=0.575, angle=0]{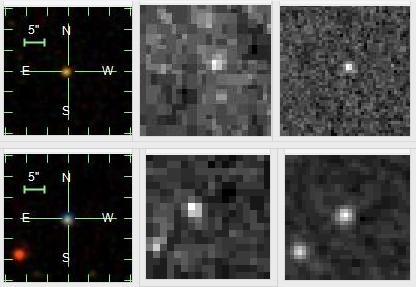} 
\caption{Two of the 11 moving stars in DR16Q, all images have 30-arcsecond sides and are centred on the SDSS location.  Top row: J012808.64+005247.1, left is the SDSS image, centre is a 1950s-epoch POSS-I image from the Digitized Sky Survey showing the star position then, right is a 1980s-epoch POSS-II image of an intermediate star position; Gaia-DR2 gives this object as having proper motion of 73 mas/yr with bearing of 101$^{\circ}$ E of N.  Bottom row: J123832.52+584022.0, images as above, Gaia-DR2 gives proper motion of 90 mas/yr with bearing of 245$^{\circ}$ E of N.} 
\end{figure}

\subsection{Processing duplicates}    

The five processing duplicates:  the DR16Q paper Section 3.1 describes its uptake of earlier visually-confirmed catalogues as ``\textit{DR7Q and DR12Q were coordinate-matched with a 0.5" radius to the superset, and missing objects were added}''.  This procedure missed 2 cases where the DR7Q quasar coordinates are offset more than 0.5 arcsec from the DR16Q coordinates, thus creating duplicate entries in DR16Q; they are J025051.30-072449.5 \& J025051.31-072450.1 (DR7Q) at an offset of 0.721 arcsec, and J093024.32+525255.7 \& J093024.29+525256.2 (DR7Q) at an offset of 0.595 arcsec.  The two resultant duplicates are listed as rows 64 \& 65 in Table 2.  Two blank-sky duplicates (J084736.12+445044.8 and J134554.61+554435.8) appear to be photometric reduction artefacts close to the true quasars, with both pairs reported in DR16Q masquerading as close quasar pairs; they are rows 66 \& 67 in Table 2.  And one blank-sky duplicate (J022613.77+035835.8) is astrometrically offset 1.4 arcseconds from the true quasar, evidently because it was fixed onto a phantom image\footnote{to view this spectrum, look up this object on the SDSS DR16 Finding Chart Tool (http://skyserver.sdss.org/dr16/en/tools/chart/chartinfo.aspx) and then select ``All spectra'' from the left bar to get access to the spectrum -- the 2-arcsecond rule prevents it from being shown as a default, because it is not a ``primary'' spectrum -- the other image has the primary spectrum.} from a double-exposed imaging run.  Both images are reported in DR16Q as separate quasars, the duplicate is row 68 in Table 2.

\section{LINE POACHERS -- false quasars}

Line poachers are stars or galaxies wrongly identified as quasars because their spectra are contaminated by stray light from nearby true quasars, in spite of any deblending efforts; thus their spectra show phantom emission lines ``poached'' from the neighbouring quasars.  This classification hazard may be worse for visual classifiers finely-tuned to recognizing quasar lines than for the pipeline; indeed, 9 of the following 12 line poachers were visually classified with high confidence (so flagged) and their assessment of the redshift was fully accurate, but not for that object.  Characteristic of line poaching is that the poacher has a very different photometric profile and appearance compared with the nearby quasar, whereas quasar lensing or NIQs (near identical quasars) generally feature similar-looking sources.  Line-poacher spectral line echoes always have a lesser flux than that of the true emission lines of the nearby quasar, see their y-axes to compare the fluxes. 
  
Of course most line poachers are assigned the same redshift as that of the neighbouring true quasar, but redshifts can be measured to be different if lines are identified differently, for example if a quasar Ly$\alpha$ line is identified as a Mg line for the line poacher.  Also it is possible that a particular line poacher gets its spectral lines not from spectral contamination but because it contains a faint lensed image of the true quasar (as better resolution might show), in which case it would be better characterized as a lens; I nominate 2 of the poachers as lensing candidates in Section 5.3.  And at the end I present a line poacher onto a pipeline quasar, with no visual inspection of either object, in which the poacher not only takes the redshift of the quasar but also takes its spectrum entirely, with the true quasar sadly lost.

\begin{figure} 
\includegraphics[scale=0.43, angle=0]{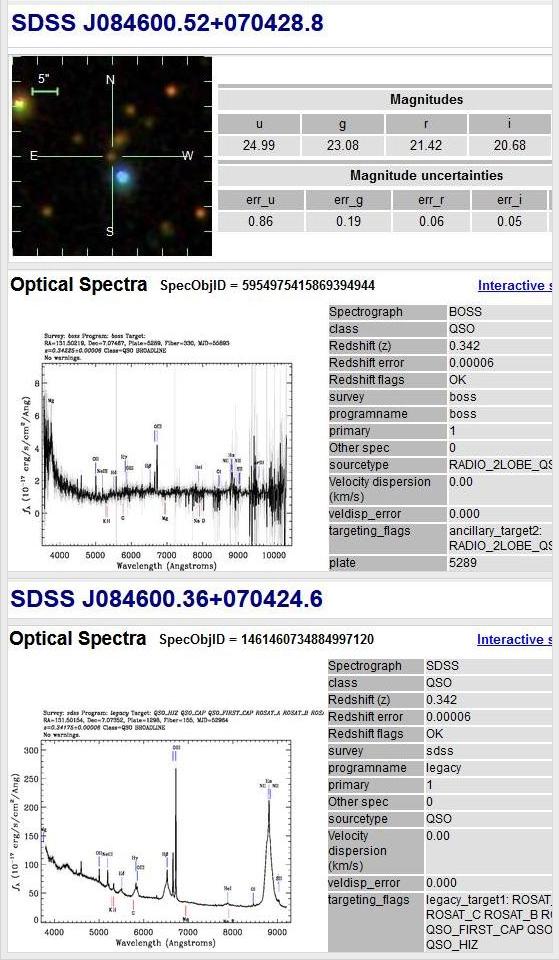} 
\caption{Line poacher: poacher ID, image \& spectrum at top, quasar ID \& spectrum at bottom.  Image shows a faint yellow star at centre, offset 4.793 arcsec from the bright blue quasar HS 0843+0715, z=0.342.  Emission lines and continuum of the quasar have contaminated the spectrum of the star, causing the pipeline to assign it the same redshift as the quasar.} 
\end{figure}

\begin{figure} 
\includegraphics[scale=0.43, angle=0]{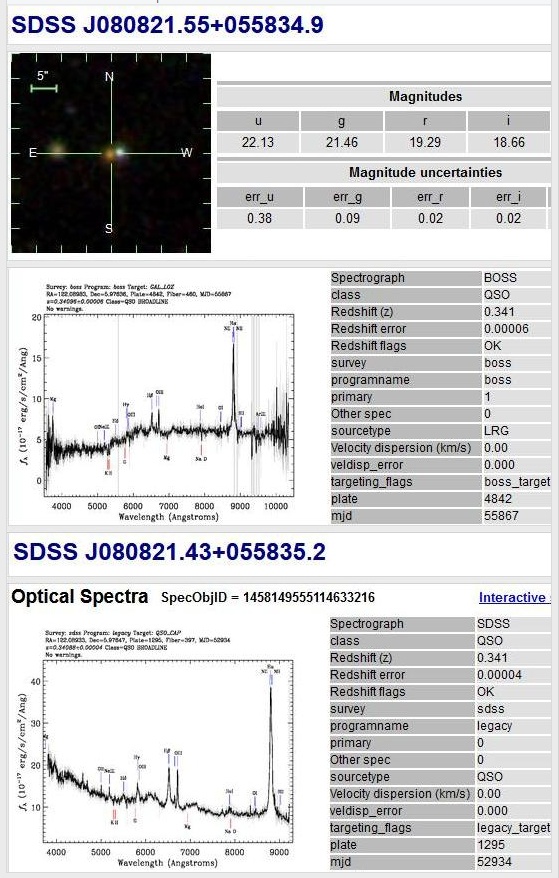} 
\caption{Line poacher: a yellow star showing emission lines of a z=0.341 quasar; at bottom, the spectrum of the true quasar.  Separation between them is 1.825 arcsec.  Emission lines of the quasar have contaminated the spectrum of the yellow star.} 
\end{figure}

\subsection{A most obvious line poacher}

Figure 17 shows a most obvious example of a line poacher, see caption for layout of this and following figures, zoom in to see clearly.  The emission lines and blueward flux of a bright quasar are echoed in the spectrum of a faint yellow star.  The star is at centre of the image, the bright blue quasar is offset 4.793 arcseconds to its lower right (south-west).  Looking at the image of the star, it is seen to be a typical member of a faint yellow star cluster there, which is a strong indicator of its classification because if it were not a star, then it would be quite coincidental that it so closely resembles those stars.  The quasar spectrum (at bottom) shows strong OIII and H$\beta$ lines at centre, and H$\alpha$ and NII lines at right, and those lines are faintly echoed in the star spectrum, causing the pipeline to assign it the same redshift as the quasar.  Also, the star's spectral flux increases at the blueward end, entirely incompatible with its faint \textit{u} and \textit{g} photometry (given at right of image), so that blue flux is clearly just further contamination from the quasar.  Thus the faint star, classified as a quasar in DR16Q, is seen to be not a quasar.  This example shows that ``line poaching'' needs to be considered when classifying objects.

\subsection{Ordinary line poachers}

Figure 18 shows the copy-over of the quasar emission lines from a white quasar to a yellow star, similar to Figure 17.  The quasar is a legacy (DR7Q) object and the star's spectrum was visually inspected as part of the subsequent BOSS project and classified as a QSO with z=0.341 with high confidence (so flagged).  The classifier, working without the image, would probably not have been aware of the very close QSO.  Note that this doublet has a separation of 1.825 arcsec, thus within the ``2 arcsecond rule'' discussed in Section 2, and has left this quasar with no primary spectrum.  This would have prevented its uptake into DR16Q which uses only primary spectra (DR16Q paper, Section 3.5 end) except that as a DR7Q quasar it was thus fortuitously included into DR16Q as well.         

\begin{figure} 
\includegraphics[scale=0.43, angle=0]{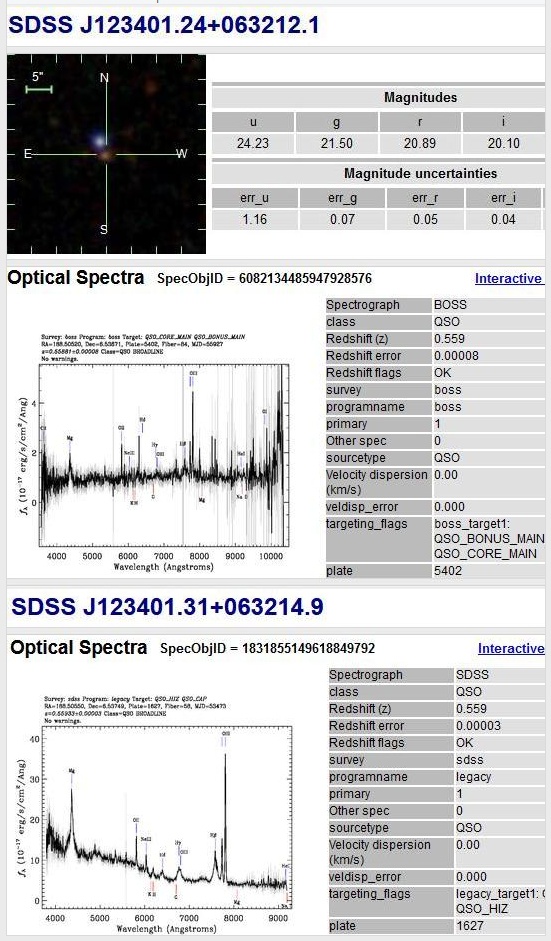} 
\caption{Line poacher: a reddish object echoes multiple strong emission lines of the mag-19 blue quasar offset 2.952 arcseconds to its upper left, thus earning the same redshift of z=0.559.  The poacher spectrum (centre) is reasonably flat into the bluest wavelengths, despite its \textit{u}=24.23 photometry, indicating continuum contamination from the quasar, as well as the lines.} 
\end{figure}

Figure 19 shows a red object's spectrum getting significant emission line contamination from the bright nearby quasar, with separation of 2.952 arcsec.  The result is that their redshifts are measured to be the same, z=0.559.  The red object is a ``bonus'' target, which means a fibre was available for this location and observers decided to target the nearby object instead of repeating the observation of the legacy quasar.  

\begin{figure} 
\includegraphics[scale=0.43, angle=0]{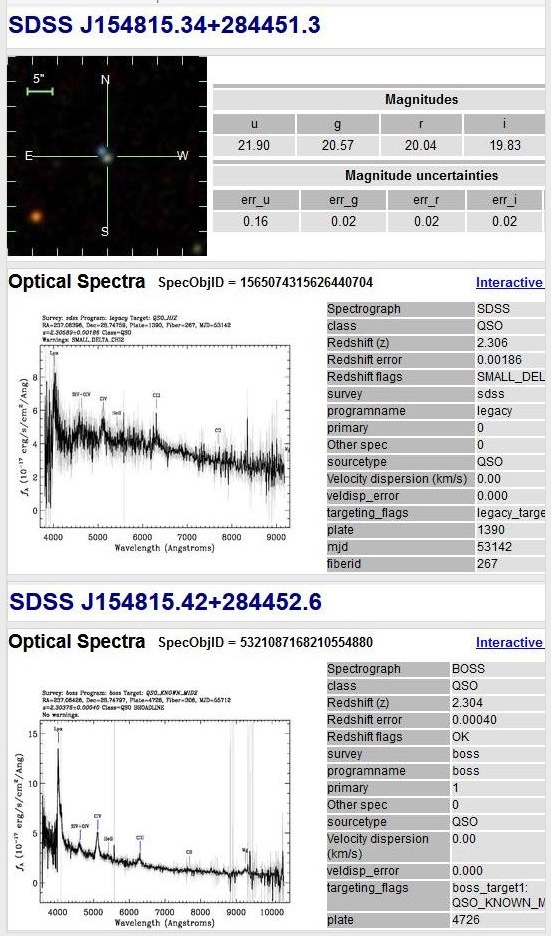} 
\caption{Line poacher: a close grey-blue doublet separated by 1.674 arcseconds.  The grey object spectrum (centre) echoes the emission lines of the blue quasar.  The echoed lines are strong, about half the flux of the quasar lines, raising the possibility of a lensed image within the grey source.} 
\end{figure} 

\begin{figure} 
\includegraphics[scale=0.43, angle=0]{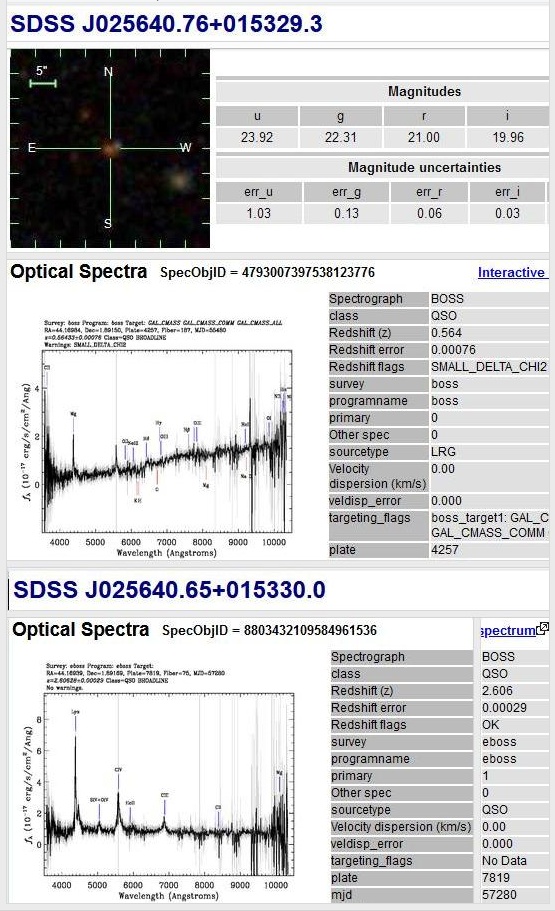} 
\caption{Line poacher: A red galaxy next to a blue quasar of z=2.606, the separation between them is 1.741 arcsec.  The galaxy is pipeline-classified as a quasar of z=0.564 due to the quasar L$\alpha$ line being taken as a Mg line, and with the quasar CIV line at 5580\AA\ being plainly visible on the galaxy spectrum but uncatered for.  That and the poor fitting of the other quasar lines shows that the pipeline redshift of 0.564 is spurious.} 
\end{figure}

Six other line poachers in Table 2 are routine, with quasar emission lines faintly echoed in the poacher spectrum.  Rows 72, 75, \& 80 of Table 2 are of ``bonus'' objects, similar to that of Figure 20, where an object near a known quasar was selected for observation; the unforeseen consequence was to get more line-poachers into the DR16Q data.  Row 76 is that of an expected faint ELG which turned out to be cleanly echoing CIV, CIII, and Mg emission lines from the 3-arcsec offset quasar J142232.53+523938.0, thus assigned with the quasar redshift of 1.49.  And rows 78 \& 79 are of reddish objects offset about 4 arcsec from low-redshift 2SLAQ \citep{2SLAQ} quasars which were not targeted by SDSS, in each case being assigned the same redshift as those neighbouring quasars.  One of those, J213615.41+000250.6, shows a bluish spectrum not compatible with the reddish photometry, raising the possibility that the spectrum pointing may have been more toward the blue quasar.  The 2SLAQ spectra are not available for comparison, but the redshifts match exactly.

\subsection{Line poachers which could be lenses}

Figure 20 shows a line poacher of an unusual grey colour, the spectrum of which includes the emission lines of its quasar neighbour -- the grey colour and the strong lines hint at a possible lensed image within the unresolved object image.  Their separation is 1.674 arcsec, thus the 2-arcsecond rule applies, so the grey object has no ``primary'' spectrum and thus displays no spectrum on its default SDSS finding chart; the user must select "All Spectra" to access that spectrum.

Figure 21 shows a red galaxy offset from a z=2.606 QSO at an offset of 1.741 arcseconds, which is within the 2-arcsecond rule so depriving the galaxy of a "primary" spectrum -- the spectrum shown is accessed by selecting the ``All Spectra'' option, as before.  The quasar emission lines are echoed on the galaxy spectrum, so this looks like an ordinary line-poaching object.  However, this has been reported as a ``confirmed quasar lens'' by the BOSS Quasar Lens Survey \citep[BQLS:][]{BQLS} which measured the galaxy redshift as z=0.603 but did not resolve the presumed quasar lens image from that of the galaxy.  The paper states ``The full imaging follow-up results will be presented in an upcoming paper'' which, however, never appeared.  The appearance of this object is that of a routine line poacher, but given the BQLS finding it is presented in this section of possible lenses.

\subsection{A quasar poached and lost}

Figure 22 shows a red star (top half of figure) line-poaching onto a mag-21 blue quasar of z=2.204; the quasar (bottom half) has a clean spectrum of strong emission lines and flat continuum; its 3 prominent Ly$\alpha$, CIV, and CIII lines are echoed on the poacher's spectrum.  The red object is classified as a quasar in DR16Q, the blue object is not; it got lost along the way.  The red object first appeared as a quasar in the DR14 pipeline and DR14Q visual catalogues, with the blue object absent.  Both objects appear and are classified as quasars in the DR16 pipeline catalogue, but the DR16Q catalogue still shows only the red object.  Their separation is 1.913 arcsec, so the 2-arcsecond rule applies and only the red object has a ``primary'' spectrum (perhaps because it was the first catalogued) -- without which the blue quasar failed to make it into DR16Q.  But the on-line SDSS-DR16 finding chart provides a further twist: when looking up the blue object, the user will find that its spectrum is neither displayed nor available from the ``All Spectra'' option, the spectrum simply isn't there at all.  In fact it is available only as a non-primary spectrum of the red object, even though the spectrum pointing co-ordinates match to the blue object, not the red -- as the reader can confirm by comparing the spectrum RA/Dec (zoom in to see those above the blue-object spectrum) to the positional decimal RA/Dec (above the blue-object image).  Somehow the 2-arcsecond rule has robbed the true quasar of its rightful spectrum, seemingly condemning it to obscurity.  However, the blue object is allocated with its spectrum and classified as a quasar in the DR16 pipeline catalogue.     

\begin{figure} 
\includegraphics[scale=0.35, angle=0]{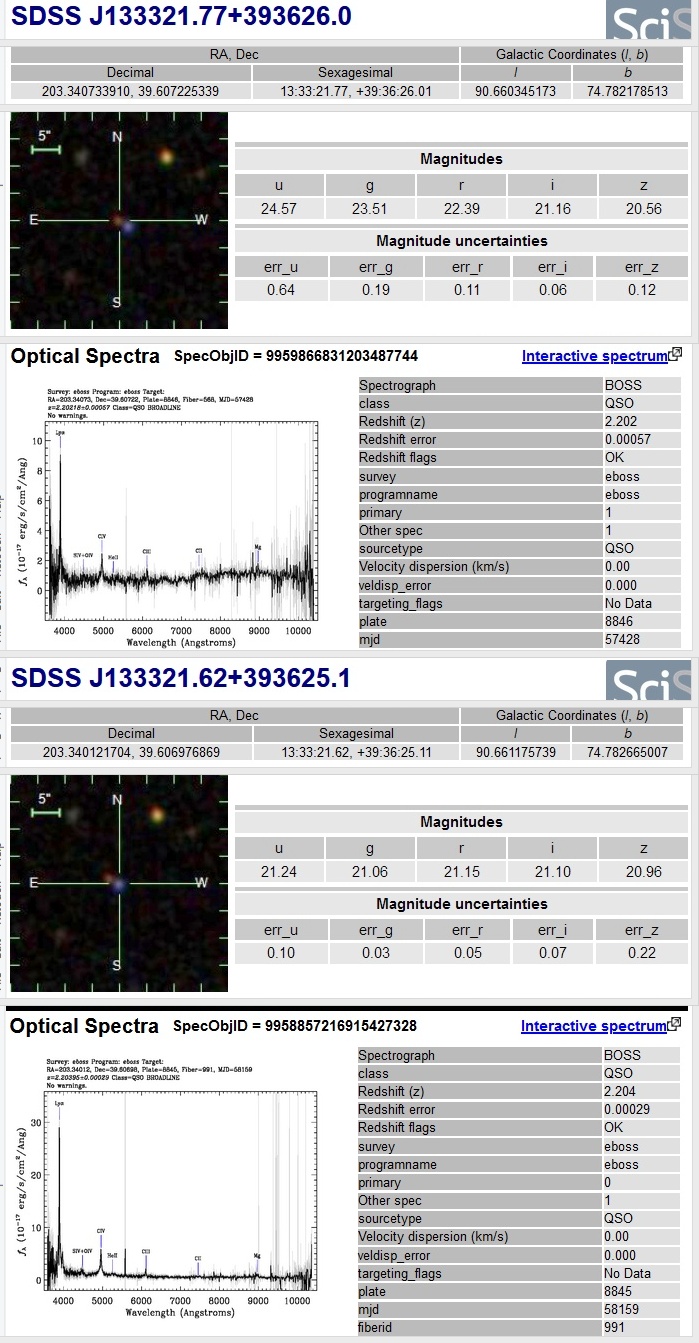} 
\caption{Line poacher: A red star offset 1.913 arcsec from a blue quasar.  The quasar spectrum at bottom shows classic lines \& flat continuum and is co-ordinate matched to the blue object, and is so reported by the DR16 pipeline, but the SDSS-DR16 finding chart allocates it to the red poacher.  The line-poacher's spectral lines are seen to exactly echo the quasar's emission lines, causing it to be assigned the quasar's redshift.  DR16Q classifies the poacher as a quasar, and does not include the true quasar.} 
\end{figure}

\subsection{Line poachers: an instrumental artefact as a new classification}  

For all of these, the line-poaching objects are classified as quasars in the DR16Q main catalogue, but wrongly so.  They are listed as lines 69 - 80 in Table 2, below. 

If a whole-sky redshift database is searched for close point-source doublets, the results will comprise 4 types: (1) quasar lenses, (2) NIQs (nearly identical quasars), (3) projected QSOs (of different redshifts), and (4) QSO / line poacher pairings.  Very little attention has been paid to the last of these, where instrumental limitations impact the classification.  Hopefully it will be more attended to in future.

\begin{table*} 
\scriptsize	 
\caption{82 non-quasars in the DR16Q Quasar catalogue}
\begin{tabular}{rllcccccl} 
\hline 
\  & type of  &                          & DR16Q & \multicolumn{3}{c}{SDSS photometry} & doublet  \\
\# & non-quasar & ID \& J2000              &   z   & \textit{u} & \textit{g} & \textit{r} & sep asec  & comment \\
\hline
1 & doublet star & SDSS J224430.44+233932.5 & 2.705 & 24.07 & 21.90 & 21.60 & 1.993 & J224430.35+233934.1 is the QSO, see Sec 2  \\
2 & doublet star & SDSS J135307.59+450854.1 & 3.035 & 23.70 & 22.24 & 21.18 & 1.843 & J135307.43+450853.6 is the QSO, see Sec 2  \\
3 & doublet star & SDSS J103745.92+061302.3 & 0.810 & 24.63 & 25.11 & 20.46 & 1.270 & J103745.85+061301.5 is the QSO, see Sec 3.0  \\
4 & doublet star & SDSS J101012.65+560520.5 & 2.136 & 22.42 & 20.77 & 19.69 & 1.036 & J101012.77+560520.0 is the QSO, see Sec 3.1  \\
5 & doublet star & SDSS J154316.78+402634.9 & 0.000 & 24.63 & 22.60 & 20.85 & 1.527 & J154316.72+402636.3 is the QSO, see Sec 3.1  \\
6 & doublet star & SDSS J093859.23+020925.8 & 1.411 & 20.30 & 20.16 & 19.76 & 1.675 & J093859.26+020924.4 is the QSO, see Sec 3.1  \\
7 & asteroid & SDSS J010858.45+053402.7 & 0.609 & 22.86 & 21.36 & 20.82 &   &    \\
8 & asteroid & SDSS J101243.07+340053.5 & 0.677 & 23.25 & 21.34 & 24.81 &   &    \\
9 & asteroid & SDSS J233240.53+044038.9 & 1.319 & 22.02 & 20.31 & 24.88 &   &    \\
10 & star spike & SDSS J000242.70+055035.7 & 1.796 & 24.63 & 23.94 & 20.76 &   &    \\
11 & star spike & SDSS J003851.42+252706.2 & 0.552 & 24.63 & 21.12 & 20.95 &   &    \\
12 & star spike & SDSS J004611.72+015954.8 & 1.518 & 24.56 & 22.58 & 21.22 &   &    \\
13 & star spike & SDSS J010800.09+210417.0 & 1.039 & 24.29 & 23.36 & 20.85 &   &    \\
14 & star spike & SDSS J011042.29+254331.5 & 4.928 & 24.63 & 24.19 & 20.90 &   &    \\
15 & star spike & SDSS J011118.84+255112.0 & 6.285 & 23.95 & 23.88 & 21.66 &   &    \\
16 & star spike & SDSS J011200.99+212245.0 & 1.007 & 24.63 & 21.73 & 21.02 &   &    \\
17 & star spike & SDSS J011700.13+281050.9 & 1.477 & 24.75 & 24.29 & 21.98 &   &    \\
18 & star spike & SDSS J013218.98+221017.5 & 0.675 & 24.67 & 22.14 & 25.53 &   &    \\
19 & star spike & SDSS J013332.74+260434.7 & 6.856 & 24.63 & 23.37 & 21.29 &   &    \\
20 & star spike & SDSS J014837.28+222437.0 & 1.091 & 24.47 & 23.71 & 21.48 &   &    \\
21 & star spike & SDSS J015830.93-041849.7 & 1.105 & 24.63 & 23.35 & 21.61 &   &    \\
22 & star spike & SDSS J015853.65+182048.4 & 6.916 & 25.87 & 23.05 & 19.62 &   &    \\
23 & star spike & SDSS J020612.24+220642.1 & 5.714 & 24.28 & 22.73 & 22.16 &   &    \\
24 & star spike & SDSS J020701.99+062044.2 & 0.768 & 24.63 & 21.99 & 22.28 &   &    \\
25 & star spike & SDSS J084617.95+195201.0 & 1.081 & 24.63 & 21.57 & 21.78 &   &    \\
26 & star spike & SDSS J084852.79+190329.4 & 0.697 & 24.63 & 22.09 & 21.94 &   &    \\
27 & star spike & SDSS J085032.25+221836.6 & 1.097 & 24.63 & 21.22 & 21.70 &   &    \\
28 & star spike & SDSS J085428.89+554717.2 & 1.769 & 24.63 & 22.94 & 21.11 &   &    \\
29 & star spike & SDSS J090827.61+372833.6 & 1.671 & 24.64 & 22.08 & 21.53 &   &    \\
30 & star spike & SDSS J093403.77+263711.0 & 5.846 & 24.56 & 21.50 & 24.80 &   &    \\
31 & star spike & SDSS J094415.42+470437.0 & 0.996 & 24.63 & 21.99 & 24.80 &   &    \\
32 & star spike & SDSS J095503.93+183659.8 & 0.496 & 24.64 & 21.46 & 21.75 &   &    \\
33 & star spike & SDSS J095953.56+534639.8 & 1.141 & 24.63 & 22.63 & 20.86 &   &    \\
34 & star spike & SDSS J105953.56+333648.9 & 0.469 & 24.64 & 23.29 & 21.10 &   &    \\
35 & star spike & SDSS J112043.12+482659.5 & 6.905 & 24.63 & 22.91 & 21.85 &   &    \\
36 & star spike & SDSS J120031.89+354629.3 & 1.704 & 24.63 & 23.34 & 21.13 &   &    \\
37 & star spike & SDSS J132007.08+462857.2 & 1.001 & 24.63 & 21.66 & 19.93 &   &    \\
38 & star spike & SDSS J145802.83+401331.1 & 2.180 & 24.63 & 20.06 & 24.80 &   &    \\
39 & star spike & SDSS J154253.66+351540.1 & 2.129 & 24.63 & 22.64 & 20.97 &   &    \\ 
40 & star spike & SDSS J155312.39+442233.3 & 2.266 & 24.90 & 21.51 & 23.31 &   &    \\
41 & star spike & SDSS J163536.05+301954.5 & 0.822 & 24.63 & 21.27 & 21.01 &   &    \\
42 & star spike & SDSS J171538.75+391133.0 & 0.466 & 24.57 & 21.25 & 20.55 &   &    \\
43 & star spike & SDSS J172627.56+352447.9 & 2.257 & 24.63 & 21.85 & 22.25 &   &    \\
44 & star spike & SDSS J224204.99+190100.5 & 5.630 & 24.63 & 23.32 & 20.74 &   &    \\
45 & star spike & SDSS J224225.15+223836.0 & 1.053 & 24.63 & 21.19 & 22.33 &   &    \\
46 & star spike & SDSS J224322.29+305354.9 & 1.692 & 24.63 & 22.19 & 24.80 &   &    \\
47 & star spike & SDSS J232505.19+205502.7 & 1.111 & 24.64 & 20.87 & 20.47 &   &    \\
48 & star glow & SDSS J000628.22+293205.1 & 0.693 & 24.63 & 21.99 & 21.31 &   &    \\
49 & star glow & SDSS J013738.20+223301.2 & 1.860 & 24.63 & 22.59 & 21.26 &   &    \\
50 & star glow & SDSS J074745.04+291706.7 & 1.771 & 24.44 & 22.00 & 25.14 &   &    \\
51 & star glow & SDSS J094303.81+450821.2 & 1.980 & 24.63 & 21.99 & 24.12 &   &    \\
52 & star glow & SDSS J111352.00+281654.1 & 0.568 & 25.03 & 24.56 & 21.29 &   &    \\
53 & moving star & SDSS J000300.27+015230.3 & 3.030 & 20.21 & 19.97 & 20.01 &   & proper motion 47 mas/yr (Gaia), DSS POSS-I confirm  \\
54 & moving star & SDSS J002958.66+283525.5 & 3.209 & 20.34 & 20.09 & 20.07 &   & proper motion 41 mas/yr (Gaia), DSS POSS-I confirm  \\
55 & moving star & SDSS J012808.64+005247.1 & 0.787 & 23.53 & 20.99 & 19.65 &   & proper motion 73 mas/yr (Gaia), DSS POSS-I confirm  \\
56 & moving star & SDSS J044911.73-063619.7 & 2.201 & 20.50 & 19.70 & 19.38 &   & proper motion 64 mas/yr (Gaia), DSS POSS-I confirm  \\
57 & moving star & SDSS J070731.45+382531.0 & 1.102 & 20.58 & 20.24 & 20.03 &   & proper motion 32 mas/yr (Gaia), DSS POSS-I confirm  \\
58 & moving star & SDSS J091106.98+541547.8 & 1.208 & 19.89 & 19.57 & 19.53 &   & proper motion 48 mas/yr (Gaia), DSS POSS-I confirm  \\
59 & moving star & SDSS J092241.95+462638.8 & 1.133 & 19.72 & 19.33 & 19.33 &   & proper motion 40 mas/yr (Gaia), DSS POSS-I confirm  \\
60 & moving star & SDSS J105525.25+620108.2 & 0.685 & 20.58 & 20.23 & 20.05 &   & proper motion 33 mas/yr (Gaia), DSS POSS-I confirm  \\
61 & moving star & SDSS J110406.68+203528.6 & 2.193 & 16.77 & 17.21 & 17.71 &   & proper motion 34 mas/yr (Gaia), DSS POSS-I confirm  \\
62 & moving star & SDSS J123832.52+584022.0 & 1.810 & 20.07 & 19.40 & 19.14 &   & proper motion 89 mas/yr (Gaia), DSS POSS-I confirm  \\
63 & moving star & SDSS J233056.81+295652.6 & 2.320 & 21.34 & 19.85 & 19.65 &   & proper motion 29 mas/yr (Gaia), DSS POSS-I confirm  \\
64 & duplicate & SDSS J025051.30-072449.5 & 2.079 & 19.37 & 19.25 & 20.95 & 0.750 & J025051.31-072450.1 from DR7Q is present  \\
65 & duplicate & SDSS J093024.32+525255.7 & 2.919 & 20.59 & 19.29 & 20.69 & 0.590 & J093024.29+525256.2 from DR7Q is present  \\
66 & duplicate & SDSS J084736.12+445044.8 & 2.782 &       &       &       & 2.073 & blank sky, 2 asec to J084736.14+445042.8, z=2.781  \\
67 & duplicate & SDSS J134554.61+554435.8 & 2.215 & 23.69 & 24.13 & 23.96 & 1.090 & blank sky, 1 asec to J134554.49+554435.2, z=2.232  \\
68 & duplicate & SDSS J022613.77+035835.8 & 1.998 & 24.62 & 19.68 & 19.97 & 1.397 & double-exposure of J022613.78+035837.2, z=1.994 \\
69 & line poacher & SDSS J025640.76+015329.3 & 0.565 & 24.31 & 22.99 & 21.86 & 1.741 & QSO is SDSS J025640.65+015330.0, z=2.606  \\
70 & line poacher & SDSS J080821.55+055834.9 & 0.341 & 23.07 & 22.40 & 20.35 & 1.825 & QSO is SDSS J080821.43+055835.2, z=0.341  \\
71 & line poacher & SDSS J084600.52+070428.8 & 0.342 & 24.99 & 23.15 & 21.58 & 4.793 & QSO is HS 0843+0715, z=0.342  \\
72 & line poacher & SDSS J103724.07+481955.5 & 0.849 & 23.02 & 21.95 & 21.94 & 4.696 & QSO is SDSS J103724.15+481950.9, z=0.847  \\
73 & line poacher & SDSS J123401.24+063212.1 & 0.559 & 24.12 & 21.50 & 21.00 & 2.952 & QSO is SDSS J123401.31+063214.9, z=0.559  \\
74 & line poacher & SDSS J133321.77+393626.0 & 2.202 & 24.57 & 23.51 & 22.39 & 1.913 & QSO is SDSS J133321.62+393625.1, z=2.204 pipe \\
75 & line poacher & SDSS J134813.08+335202.7 & 1.738 & 23.12 & 21.45 & 20.90 & 3.393 & QSO is SDSS J134812.97+335159.6, z=1.738  \\
76 & line poacher & SDSS J142232.77+523940.1 & 1.493 & 21.58 & 21.72 & 21.62 & 3.051 & QSO is SDSS J142232.53+523938.0, z=1.495  \\
77 & line poacher & SDSS J154815.34+284451.3 & 2.304 & 21.89 & 20.56 & 20.05 & 1.674 & QSO is SDSS J154815.42+284452.6, z=2.304  \\
78 & line poacher & SDSS J212112.13-003300.2 & 0.460 & 22.06 & 21.35 & 20.76 & 4.166 & QSO is 2SLAQ J212112.04-003304.1, z=0.460  \\
79 & line poacher & SDSS J213615.41+000250.6 & 0.593 & 22.30 & 21.62 & 21.11 & 3.868 & QSO is 2SLAQ J213615.62+000252.9, z=0.593  \\
80 & line poacher & SDSS J230251.92+024915.0 & 1.417 & 22.54 & 21.89 & 21.76 & 3.018 & QSO is SDSS J230251.93+024918.0, z=1.413  \\
81 & unplugged & SDSS J110918.07+362737.0 & 2.310 & 23.33 & 23.08 & 21.32 &   & a galaxy, z=0.623, DR16Q used UNPLUGGED spectrum   \\
82 & in galaxy & SDSS J102039.30+284039.7 & 1.564 & 22.38 & 21.91 & 22.19 &   & a galaxy, DR16Q used non-primary spectrum  \\

\hline
\end{tabular}
\end{table*}

\section{Conclusion} 

40 new quasars, not previously published, are found in a search through the extensive SDSS-DR16 pipeline and SDSS-DR16Q Superset data, and are presented here.  Also 82 SDSS-DR16Q quasars are shown to be not quasars.  A key topic in both results is that spectra can be contaminated by the emission of close neighbours on the sky due to imperfect deblending, such that the classifier can be confused into giving the classification of the neighbouring object instead of the targeted object.  Thus more attention should be paid to close neighbours on the sky, when classifying objects.

\section*{Acknowledgements}
Thanks to Adam Myers for helpful discussions.  This work was not funded.

\section*{Data Availability}

All data presented by this paper is within this document; there is no external data.








\bsp	
\label{lastpage}
\end{document}